\def \cm{~\rm{cm}}
\def \s{~\rm{s}}
\def \km{~\rm{km}}

\def \K{~\rm{K}}
\def \g{~\rm{g}}

\def \yr{~\rm{yr}}

\def \kpc{~\rm{kpc}}

\documentclass[12pt,preprint]{aastex}
\usepackage{graphics,epsf}
\usepackage{amsmath}                
\usepackage{amsfonts}               
\usepackage{amssymb}                
\usepackage{epsfig}            
\begin{document}

\title{Shaping Planetary Nebulae by Light Jets}

\author{Muhammad Akashi and Noam Soker}
\affil{Department of Physics, Technion$-$Israel Institute of
Technology, Haifa 32000, Israel
akashi@physics.technion.ac.il; soker@physics.technion.ac.il}

\begin{abstract}

We conduct numerical simulations of axisymmetrical jets expanding
into a spherical AGB slow wind. The three-dimensional flow is simulated with an
axially symmetric numerical code.
We concentrate on jets that are active for a relatively short time.
Our results strengthen other studies that show that jets can account for many
morphological features observed in planetary nebulae (PNs).
Our main results are as follows.
(1) With a single jet's launching episode we can reproduce a lobe structure
having a `front-lobe', i.e., a small bulge on the front of the main lobe, such as that
in the PN Mz~3.
(2) In some runs dense clumps are formed along the symmetry axis, such as those
observed in the pre-PN M1-92.
(3)  The mass loss history of the slow wind has a profound influence on the PN structure.
(4) A dense expanding torus (ring; disk) is formed in most of our runs. The torus is
formed from the inflated lobes, and not from a separate equatorial mass loss episode.
(5) The torus and lobes are formed at the same time and from the same mass loss
rate episode. However, when the slow wind density is steep enough, the ratio of the
distance divided by the radial velocity is larger for regions closer to the equatorial
plane than for regions closer to the symmetry axis.
(6) With the short jet-active phase a linear relation
between distance and expansion velocity is obtained in many cases.
(7) Regions at the front of the lobe are moving sufficiently fast to excite some visible emission lines.
\end{abstract}


\section{INTRODUCTION}
\label{sec:intro}

The understanding of planetary nebulae (PNs) shaping is the prime
focus of the PN community for about two decades (Balick \& Frank
2002 and references therein). Some popular mechanisms for shaping
involve jets that are launched by the central star or its
companion (e.g., Morris 1987, 1990; Soker 1990, 1992, 1996; Soker \& Livio
1994; Sahai \& Trauger 1998; Soker \& Rappaport 2000; see Soker \&
Bisker 2006 for many more references and a discussion of the
development of the jet shaping models). The shaping jets in these models
are blown during the late AGB phase or early post-AGB phase,
because mass must be supplied to a companion outside the
envelope (Morris 1987, 1990; Soker \& Rappaport 2000), or in a post-common
envelope evolution (Soker \& Livio 1994). Even jets with small amount of mass
that might form small bullets (termed \emph{ansae} or FLIERS) are formed
in the pre-PN phase (Soker 1990).
Recent studies strongly support models where the main shaping occurs in
the pre-PN phase or somewhat earlier (e.g., Sahai et al. 2007).

Based on many other astrophysical objects, it can be safely
deduced that the jets are blown by accretion disks (Soker \& Livio
1994). A single AGB star does not possess enough angular momentum
to form an accretion disk, and for that the binary mechanism was
introduced (Morris 1987, 1990; Soker \& Livio 1994). In the binary
model for the formation of bipolar PNs$-$those with large lobes
and with a waist between the lobes$-$a mass transfer process from the AGB
or post-AGB star to a compact companion forms an accretion disk
around the companion (Soker 2002). The companion then blows the two opposite jets.

Hydrodynamical numerical simulations show that jets launched by the central
star (or its companion) can indeed account for many features of asymmetrical PNs
(Soker 1990; Lee \& Sahai 2003, 2004;  Garcia-Arredondo \& Frank 2004;
Velazquez 2004, 2007; Riera et al. 2005; Raga et al. 2007; Stute \& Sahai 2007;
Akashi 2008; Akashi et al. 2008; Akashi \& Soker 2008; Dennis et al. 2008;
Guerrero et al. 2008). Our main new addition is the usage of jets with wide
opening angle (Akashi 2008; Akashi \& Soker 2008; Akashi et al. 2008), that
might be more appropriately called CFW, for collimated fast wind.
In the present paper we explore more properties of lobes inflated by wide jets.
We note that rapidly precessing jets have the same effect as wide jets
(Sternberg  \& Soker 2007), but for numerical reasons we simulate wide jets.
We simulate cases where the jets is active for a relatively short time,
with the goal of reproducing the basic type of structures represented
by the PN MZ~3 (PN G331.7-01.0; `The Ant') and pre-PN M1-92
(PN G064.09+04.26; IRAS 19343+2926; `Minkowski's footprint').
The short jet-activity phase leads to a more or less ballistic evolution after
the jet is shut off, that forms a general linear relation between the distance
of gas parcel from the center and its outward velocity, as is observed in some PNs
(e.g., Mz~3, Santander-Garcia et al. 2004;
M1-92, Alcolea et al. 2007; NGC 6302, Meaburn et al. 2008).
(This linear relation is wrongly termed ``Hubble type flow''. However,
in the Hubble flow the space itself expands with no center. In a PN there
is a well defined center of the expanding gas; see Trimble et al. 2007.)

Because we do not include the ionizing radiation from the central star, we compare our results
with young PNs, or pre-PNs: The dynamical age of the pre-PN M1-92 is $\sim 1200 \yr$
(Alcolea et al. 2008), and the age of the lobes of Mz~3 is $\sim 900 \yr$ (Guerrero et al. 2004;
using a distance of 1.4~kpc from Chesneau et al. 2007).
The young age of the lobes of these objects is one of the reasons we use them for
comparison with our results.

\section{THE NUMERICAL SCHEME}
\label{sec:numerical}

The simulations were performed using Virginia Hydrodynamics-I
(VH-1), a high resolution multidimensional astrophysical
hydrodynamics code developed by John Blondin and co-workers
(Blondin et al. 1990; Stevens et al., 1992; Blondin 1994). We have
added radiative cooling to the code at all temperatures $T > 100 \K$.
Radiative cooling is carefully treated near contact
discontinuities, to prevent large temperature gradients
from causing unphysical results. The cooling function $\Lambda (T)$
for solar abundances that we use was taken
from Sutherland \& Dopita (1993; their table 6).

We simulate three-dimensional axisymmetrical morphologies. This
allows us to use two-dimensional axisymmetrical grid, and to
simulate one quarter of the meridional plane. There are 208 grid
points in the azimuthal ($\theta$) direction of this one quarter
with an equal azimuthal spacing, and 208 grid points in the radial
direction. The radial size of the grid points increases with
radius by a factor of $1.015$ from one cell to the next.
The cells' radial size at the inner boundary is $\Delta r(r_{\rm in}) =2.8 \times 10^{14} \cm$,
while it is $\Delta r(r_{\rm out}) = 6.1 \times 10^{15} \cm$ at the outer boundary.
In these simulations the grid extends from $r_{\rm in}=10^{15}
\cm$ to $r_{\rm out}=4 \times 10^{17} \cm$.

Before the CFW (jet) is launched at $t=0$ the grid is filled with a
spherically-symmetric slow wind having a uniform radial velocity of
$v(r,\theta)=v_1=10 \km \s^{-1}$.
The density at $t=0$ is taken to be
\begin{eqnarray}
\rho(t=0) = \left\{ \begin{array}{cl}
\frac {\dot M_1}{4 \pi r^{2}v_1} & \qquad 0 \le r < R \\
\frac {\dot M_1}{4 \pi R^{2}v_1} \left(\frac{R}{r}\right)^{\beta} & \qquad r > R
\end{array}
 \right .
 \label{dens}
\end{eqnarray}
where $R$ is a radius that separates two regions: an inner one where
the density varies as $r^{-2}$,  and an outer one where the density decreases
as $r^{-\beta}$. The values of $\beta$ for the different runs are shown in Table 1.
In all new runs presented in this paper we take $R=3 \times 10^{16} \cm$ and $\dot M_1=10^{-6} M_\odot \yr^{-1}$.
This density profile represents an AGB mass loss history where the mass loss rate increased
until a time $R/v_1$ before the onset of the CFW (jet), and stayed constant at a value of
$\dot M_1$ thereafter until $t=0$.

The CFW (jet) is launched from the first 20 radial cell-rows attached to the inner boundary
of the grid, extending in the range $10^{15} \cm \le r \le 6.5\times 10^{15} \cm$,
and within an angle (half opening angle) $\alpha$ ($ 0 \le \theta \le \alpha$).
In the present paper we take $\alpha=30^\circ$ and CFW initial radial velocity
of $v_2=600 \km \s^{-1}$ in all runs.
For numerical reasons a weak slow wind is injected in the sector $ \alpha < \theta \le 90^\circ$.
Here $\theta =0$ is along the symmetry axis (vertical in the figures).

The initial temperature of the slow wind in this work is $T_1=1000 \K$.
The temperature of the gas injected into the CFW is $T_2= 1000 \K$, but
it has no influence on the results because of the high mach number.
The initial temperature of the slow wind has a noticeable
influences on the final shape of the PN (Akashi 2008).
An initial temperature of $T_1=1000 \K$ is appropriate for most cases.
An initial temperature of $T_1=10^4 \K$ might be appropriate
for some cases that the companion to the AGB (or post-AGB) star is a WD.
If the companion WD accretes at a high rate it can sustain a constant nuclear
burning and becomes super-soft X-ray sources (e.g., Starrfield et al. 2004),
and ionize the nebula. This is the case for some symbiotic nebulae.

Another important parameter is the duration of the jet active
phase. The CFW (jet) is blown with unchanged parameters for a time
of $\Delta t_2$, and then turned off. The main motivation for the
short duration of the jet active phase is the observed linear
relation of velocity with distance from the center (e.g.,
Santander-Garcia et al. 2004; Meaburn et al. 2008).

Our calculations do not include the ionizing radiation and fast wind blown by the
central star during the PN phase. We are studying the shaping that occur in the pre-PN
(proto-PN) phase. Both the fast wind and the ionizing radiation play a role in
the later shaping of the nebula (Perinotto et al. 2004; Sch\"onberner et al. 2004),
but the basic shaping occurs already in the pre-PN or in the late AGB phase
(e.g., Sahai et al. 2007).

\section{THE PHYSICAL PARAMETERS}
\label{sec:param}
\subsection{The structures to be reproduced}
\label{sec:struc}

\begin{figure}
\resizebox{0.9\textwidth}{!}{\includegraphics{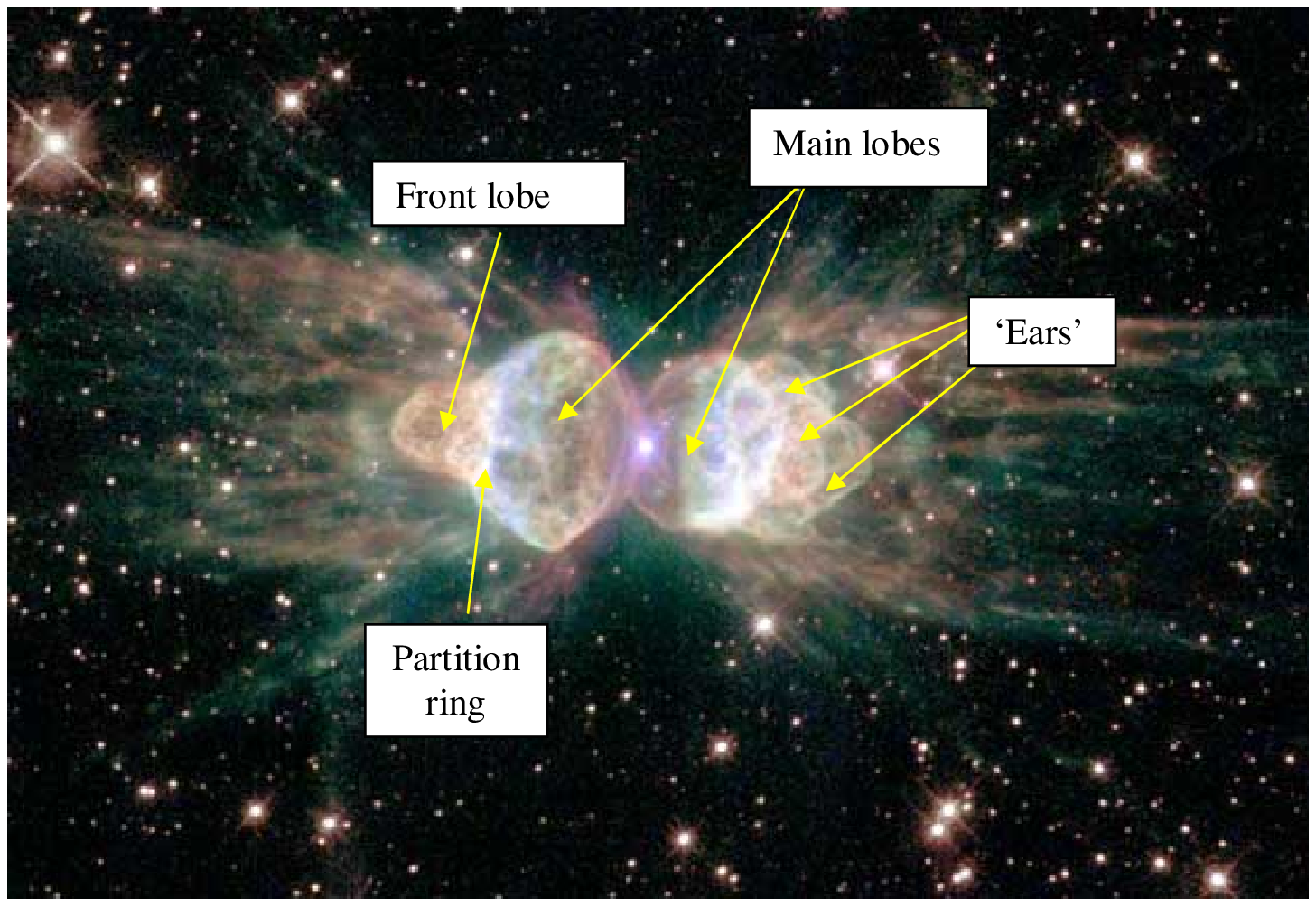}}
\centering
\caption{HST image of the Mz~3 planetary nebula
(Credit: R. Sahai, B. Balick, Hubble Heritage Team, ESA, NASA).
We mark several morphological features that we will compare with numerical results. }
\label{Mz3}
\end{figure}

Our goal is not to explain all features of a particular PN, but rather to
show that we can reproduce two typical types of bipolar structures as
represented by two PNs.
The first Structure is the one represented by Mz~3 (PN G331.7-01.0; `The Ant'),
a well studied PN (e.g., Guerrero et al. 2004).
Although Mz~3 is commonly referred to as a PN, it might actually be a
symbiotic nebula (Schmeja \& Kimeswenger  2001).
For our purpose it does not matter, as we are looking into the
general structure which is shared by many other PNs.
The image of this PN is given in Fig.~\ref{Mz3}.

Santander-Garcia et al. (2004) identify four major parts in the nebula.
(1) A pair of bright bipolar lobes. They extend to a projected distance of
$\sim 17 ^{\prime \prime}$ ($5 \times 10^{17} \cm$ at a PN distance of $D=2 \kpc$)
from the center.
(2) Two opposite highly collimated column-shaped outflows. These
are narrow features along the symmetry axis that are outside the bipolar lobes.
(3) A conical system of radial structures, also outside the bipolar lobes.
(4) A very dim, low-latitude and flattened, ring-like,  radial outflow
in the equatorial plane outside the bipolar lobes.
We will try to reproduce only the bipolar lobes, and not the structural
features outside them, that are the result of previous mass loss episodes.

There are two other structural components inside the bipolar lobes.
A silicate disk of radius $0.006^{\prime \prime}$ at the center of Mz~3
(Chesneau et al. 2007), and X-ray emitting region near the symmetry axis,
that might be related to a jet (Kastner et al.2003).
The X-ray emission and morphology suggest a collimated outflow with a
speed of several$\times 100 \km \s^{-1}$.

\begin{figure}
\resizebox{0.9\textwidth}{!}{\includegraphics{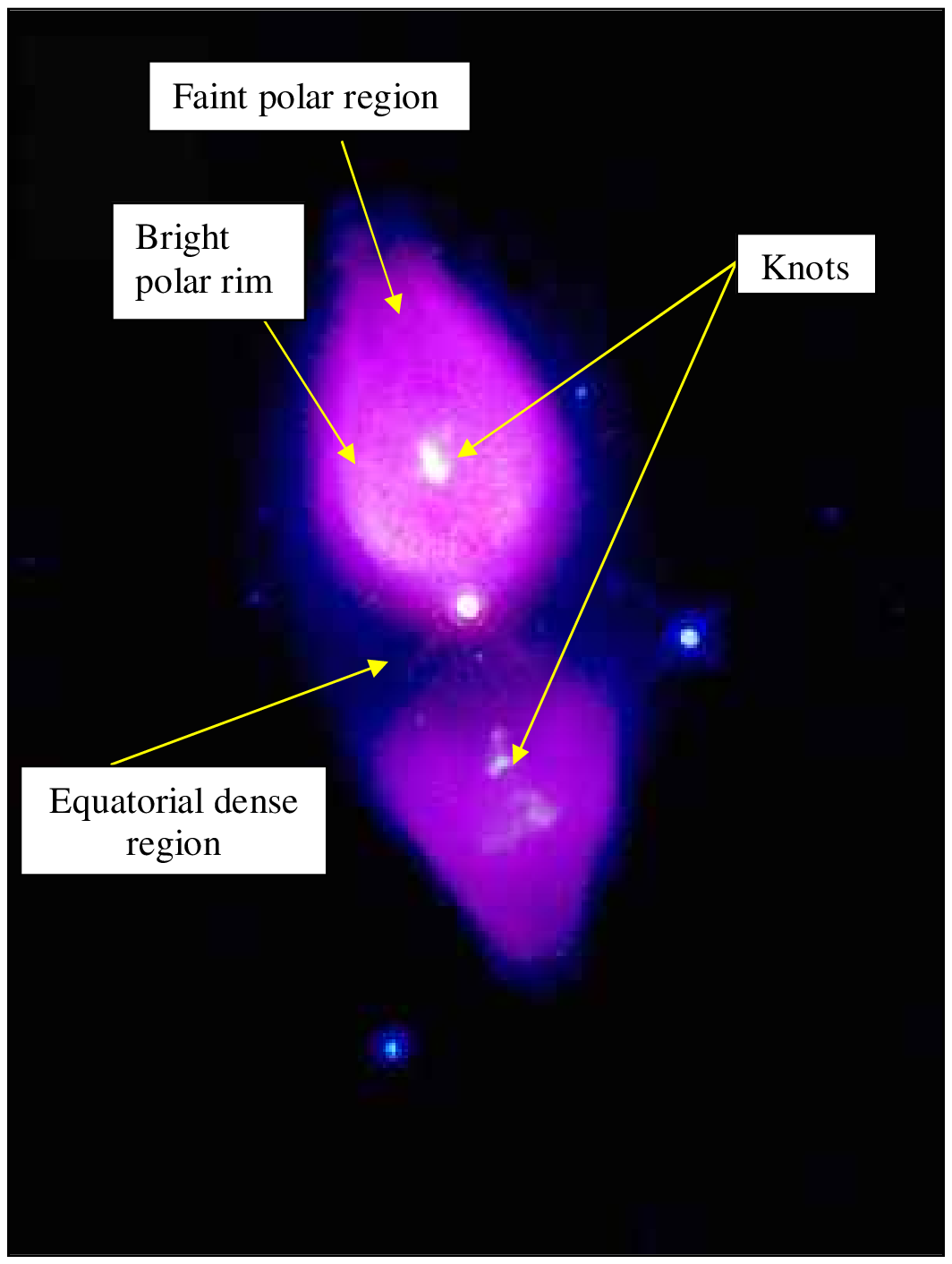}}
\centering
\caption{HST image of the M1-92 pre-planetary nebula (Credit: ESA, NASA-HST;
see Bujarrabal et al. 1998b, Trammell \& Goodrich 1996, and the Planetary Nebula Image Catalogue
maintained by B. Balick).
We mark several morphological features that we will compare with numerical results.  }
\label{M1-92}
\end{figure}

The second kind of bipolar structure we are aiming to produce is the
one represented by the bipolar pre-PN M1-92 (PN G064.09+04.26; Minkowski's footprint).
An image of this PN is given in Fig.~\ref{M1-92}.
Early studies of this pre-PN (proto-PN) were conducted by Herbig (1975),
Calvet \& Cohen (1978), and Trammell \& Goodrich (1996), and it was then extensively
studied by Bujarrabal et al. (1997, 1998a,b) and Alcolea et al. (2007, 2008).
Because it is a pre-PN, most of the mass is not observed in the visible band.
The bipolar structure seen in figure Fig.~\ref{M1-92} is the illuminated part
only; most of the mass resides in an expanding molecular disk in the equatorial plane.
The visible lobes extend to a distance of
$6 ^{\prime \prime}$ ($2.3 \times 10^{17} \cm$ at a PN distance of $D=2.5 \kpc$)
from the center.
There is a dense central disk (not an accretion disk) with a diameter of $10^{17} \cm$,
and the total mass of the molecular envelope is $\sim 1 M_\odot$
and the estimated age is $\sim 1000 \yr$ (Bujarrabal et al. 1998a,b; Alcolea et al. 2008).

\subsection{Previous runs}

\begin{figure}[h]
\centering
\resizebox{0.8\textwidth}{!}{\includegraphics{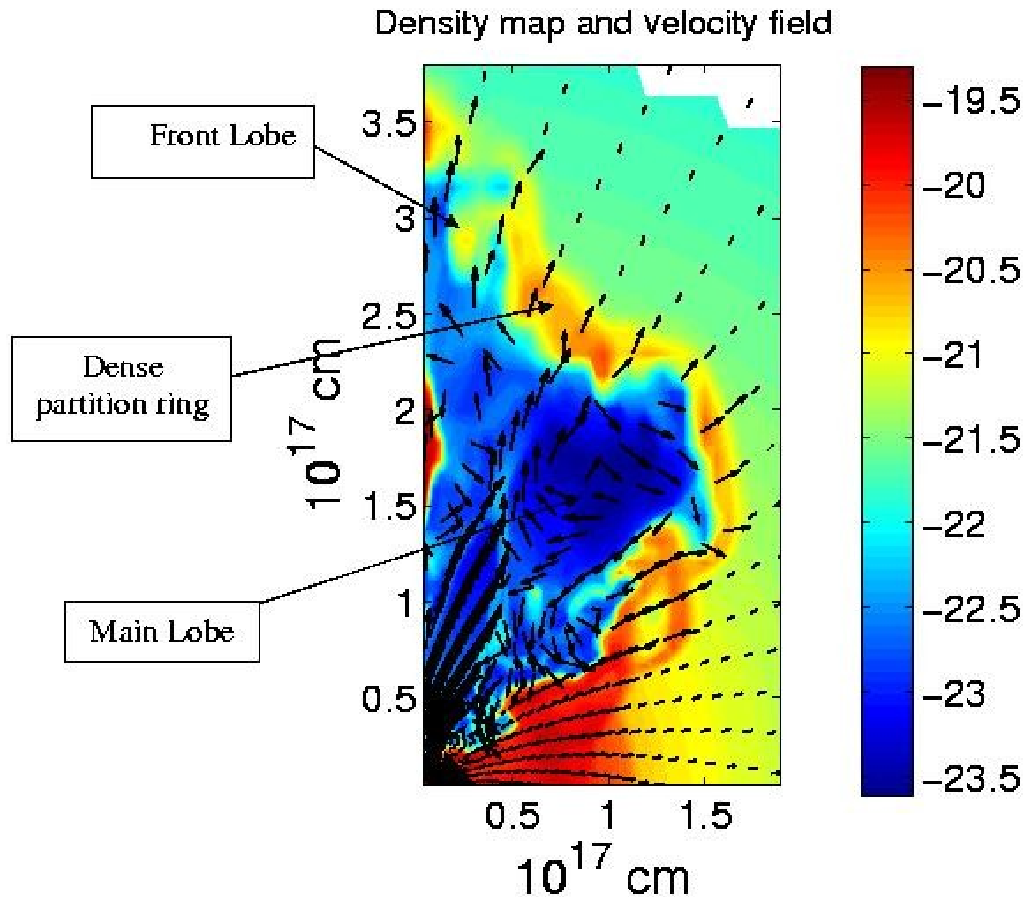}}
\hskip +2.5in
\vskip -2.9in
\centering
\resizebox{0.55\textwidth}{!}{\includegraphics{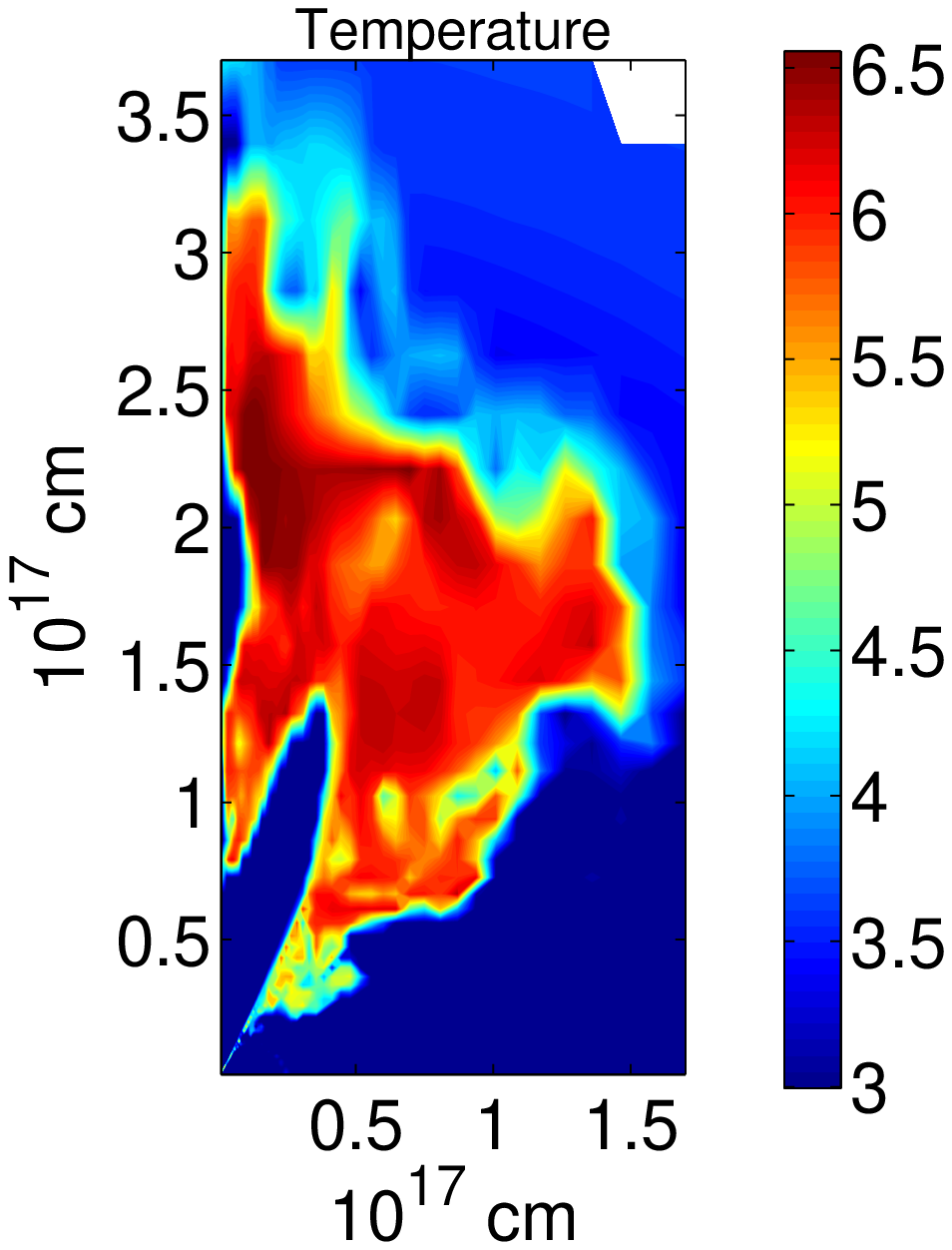}}
\caption{Results from a run conducted by Akashi (2008; it does not appear in Table 1 here).
Shown are the density and velocity map, and the temperature map, both maps at $t=1560 \yr$.
In all runs the jet (CFW) starts at $t=0$.
The initial (at $t=0$) slow wind velocity is $v_1=10 \km \s^{-1}$, and the density profile is built
for a mass loss rate of $\dot M_1 = 3 \times 10^{-5} M_\odot \yr^{-1}$ for $r < 3 \times 10^{16} \cm$,
and $\dot M_1 = 5 \times 10^{-6} M_\odot \yr^{-1}$ for $r > 3 \times 10^{16} \cm$.
The jet has a half opening angle  $\alpha=30^\circ$, it is injected with a radial velocity of
$v_2=600 \km \s^{-1}$, and a mass loss rate into one jet of $\dot M_2 = 10^{-7} M_\odot \yr^{-1}$.
The initial temperature of the slow wind and the injected jet material is
$10000 \K$, and the temperature is limited from below at $1000 \K$.
Marked are features that can be identified in the image of Mz~3 (Fig.~\ref{Mz3}).
Arrows indicate flow direction, with a scale of: $v > 200 \km \s^{-1}$ (long arrow),
$ 20 < v \le 200 \km \s^{-1}$ (medium arrows),
and $ v \le 20 \km \s^{-1}$ (short arrows).
The density scale is in logarithmic units of $\g \cm^{-3}$, and
the temperature scale is in logarithmic units of degrees in Kelvin.}
\label{M0}
\end{figure}

Before turning to describe the new runs, we present one of the previously simulated cases
(Akashi 2008), that also reproduce the basic structure of Mz~3.
In that paper different mass loss rate histories of the AGB progenitor were studied.
In a model presented here in Fig.~\ref{M0},  the slow wind mass loss rate
was increased by a factor of $k_1=6$ for a time period of $\Delta t_1=950 \yr$,
ending at the beginning of the jet-launching phase.
We marked on that figure morphological features that can be identified in the image of Mz~3
(Fig.~\ref{Mz3}).
Akashi (2008) found that the mass loss history of the AGB wind plays a significant role in
the shaping process. In the present study the mass loss history is modelled by the
density profile of the slow wind as given by equation (\ref{dens}).

\subsection{New parameters}

Aiming at the structures represented by the two PNs described in section \ref{sec:struc},
we simulated many cases, out of which we present those that are summarized in Table 1.
The first column lists the run, the second column gives the value of the mass
loss rate of one jet $\dot M_2$, the third column lists the parameter $\beta$
that appear in equation (\ref{dens}), and in the fourth column the duration of
the jet's launching (active) phase is given.
The following physical parameters were held the same for all runs:
The initial velocity of the slow wind $v_1=10 \km \s^{-1}$ and of the
CFW (jet) $v_2=600 \km \s^{-1}$, both of them radial,
the initial temperature of the slow wind and of the CFW $1000 \K$,
the radius and mass loss rate appearing in equation (\ref{dens}),
$R= 3 \times 10^{16} \cm$, and $\dot M_1=10^{-6} M_\odot \yr^{-1}$, respectively,
and the half opening angle of the jet $\alpha = 30^{\circ}$.
\begin{table}[h]

Table 1: Cases Simulated

\bigskip
\begin{tabular}{|l|c|c|c|c|c|c|c|}
\hline
Run & $\dot M_2$ & $\beta$ & $\Delta t_2$ & Figure \\
&($M_\odot \yr^{-1}$)  & &  ($\yr$) &  \\
\hline
M50 &  $10^{-7}$ & 2 & 50 & \ref{M50}, \ref{linflow2}  \\
\hline
M3-50 &  $3 \times 10^{-7}$ & 2 & 50 & \ref{M3-50} \\
\hline
M80 &   $10^{-7}$ & 2 & 80 & \ref{M80}, \ref{m2temp}, \ref{linflow1} \\
\hline
M150 &  $10^{-7}$ & 2 & 150 & \ref{M150} \\
\hline
M600 & $10^{-7}$ & 2 & 600 & \ref{M600}, \ref{m600v}  \\
\hline
M2.2 & $10^{-7}$ & 2.2 & 150  & \ref{linflow1} \\
\hline
M3 & $10^{-7}$ &  3 & 150  & \ref{M3}, \ref{compare}, \ref{tau10}, \ref{linflow2} \\
\hline
M2.2-3 & $3 \times 10^{-7}$  & 2.2 & 150  & \ref{M2.2-3} \\
\hline

\end{tabular}

\footnotesize
\bigskip
The first column gives the name of the run, the second column lists the mass loss rate of
one side of the CFW (one jet), $\beta$ is the density-profile parameter
in equation (\ref{dens}), and the 4th column gives the duration of the CFW injection
(jet launching) period.
In the last column we list the figures where results of the runs are described.
In all runs the slow wind velocity is $v_1=10 \km \s^{-1}$, the mass loss rate
and radius that appear in equation (\ref{dens}) are $\dot M_1 = 10^{-6} M_\odot \yr^{-1}$
and $R= 3 \times 10^{16} \cm$, respectively,
the initial temperature of the slow wind and the injected CFW is $1000 \K$,
the initial velocity of the CFW is $v_2=600 \km \s^{-1}$, and
the half opening angle of the CFW is $\alpha=30^\circ$.
\end{table}

\section{RESULTS AND DISCUSSION}
\label{sec:results}

\subsection{Dependance on the jet active phase duration}
We run several cases with varying jet-active period $\Delta t_2$.
One of the motivations for a short jet activity period comes from the
ballistic motion observe in some PNs, i.e., a linear relation between the
distance form the center and the radial velocity
(e.g., Santander et al. 2004;  Meaburn et al. 2008).

\begin{figure}
\resizebox{0.5\textwidth}{!}{\includegraphics{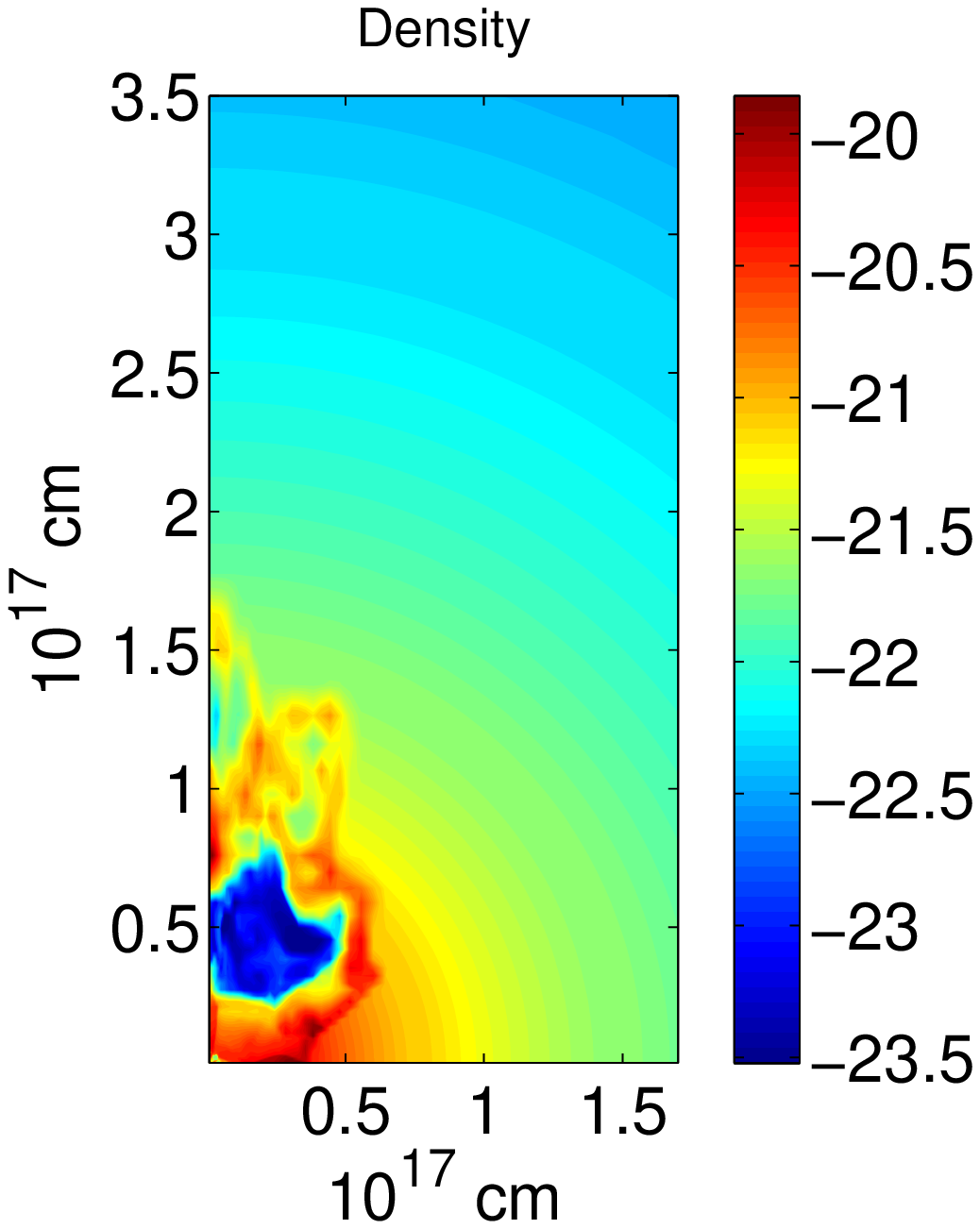}}
\hskip -1.in
\resizebox{0.5\textwidth}{!}{\includegraphics{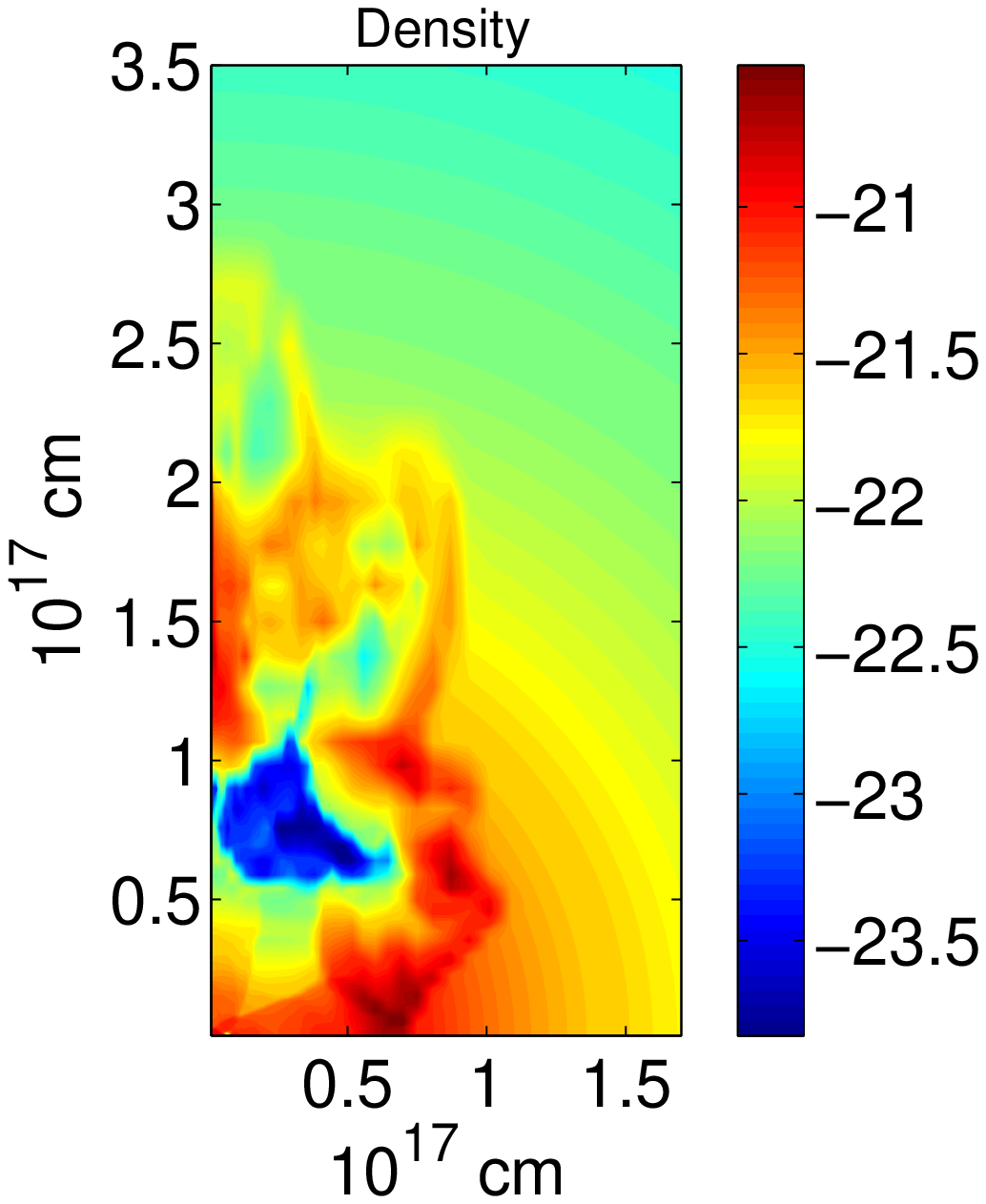}}
\caption{Density map for the M50 run at $t=540 \yr$ (left)
and $t=1140 \yr$ (right). The time is set to $t=0$ when the jet starts.
The density scale here and in all figures is in logarithmic units of
$\g \cm^{-3}$.}
\label{M50}
\end{figure}

\begin{figure}

\resizebox{0.5\textwidth}{!}{\includegraphics{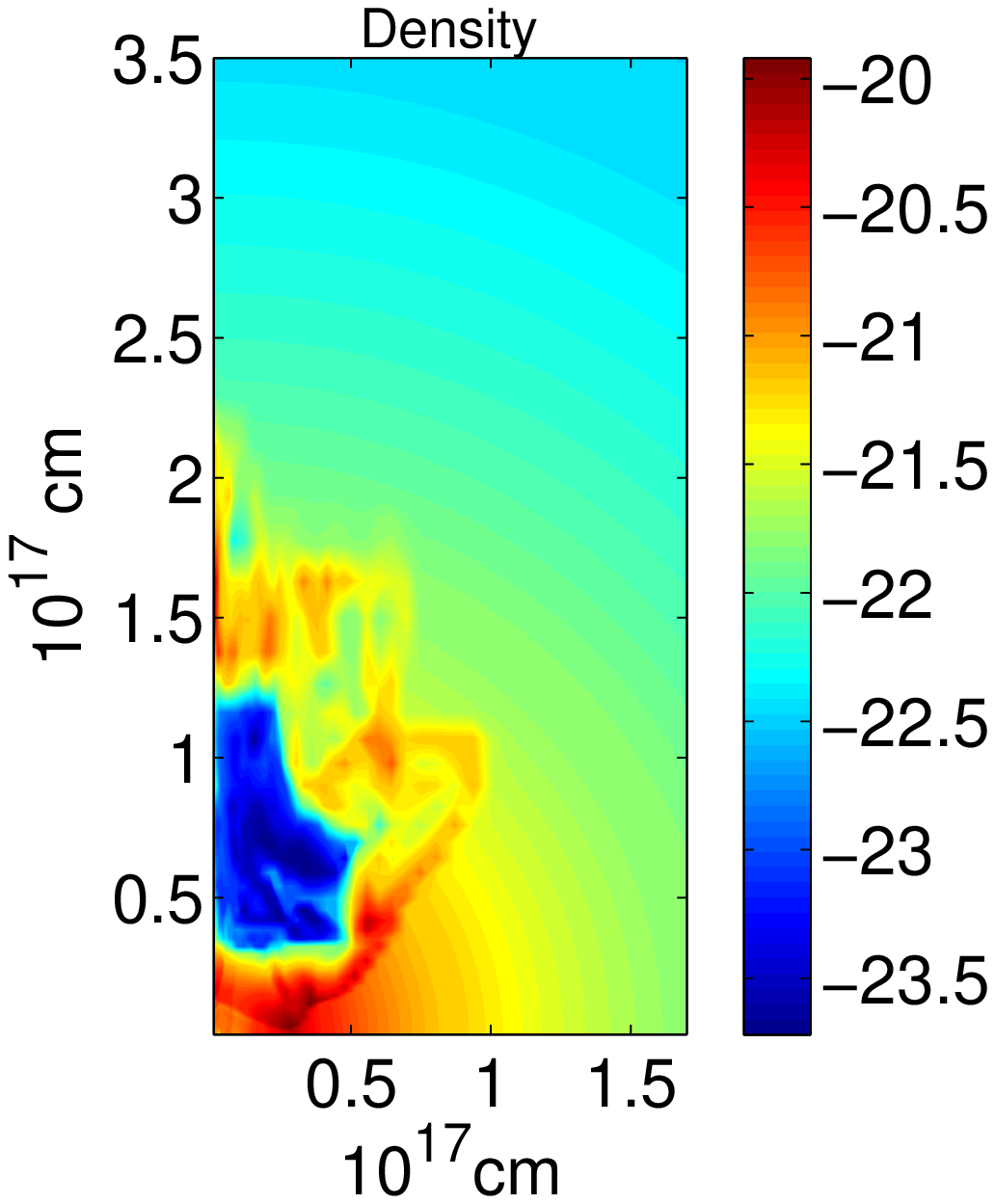}}
\hskip -1.in
\resizebox{0.5\textwidth}{!}{\includegraphics{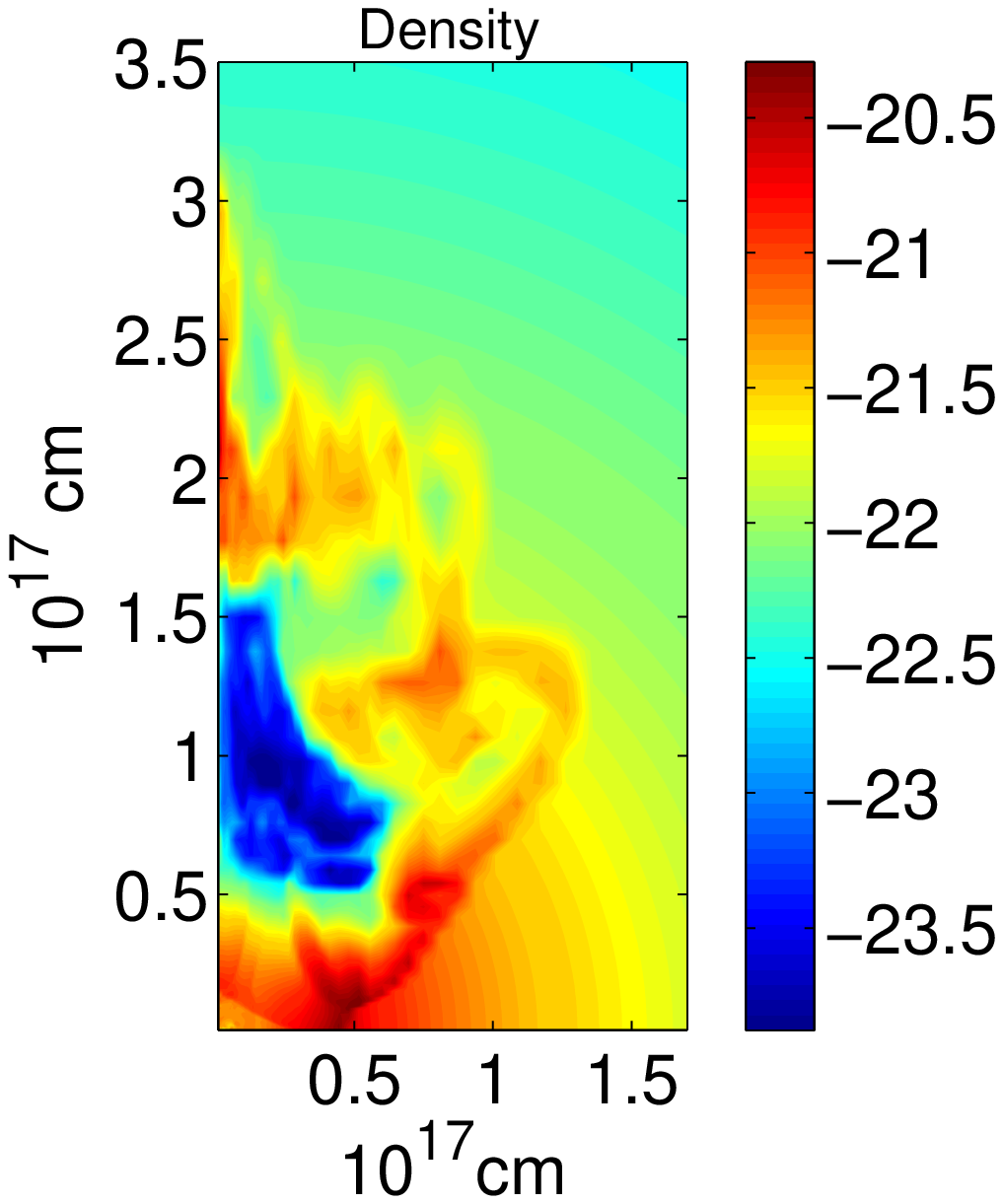}}
\caption{Density map for the M3-50 run at $t=540 \yr$ (left)
and $t=840 \yr$ (right). }
\label{M3-50}
\end{figure}

We first compare two runs differing only by the mass loss rate of the jet $\dot M_2$.
These are run M50 presented in Fig.~\ref{M50} and run M3-50 presented in Fig.~\ref{M3-50},
given each at two times as mentioned in the respective figure captions.
Run M3-50 evolves faster as more energy was injected via the jet.
This type of interaction has some instabilities that develop into `ears' protruding from
the hot bubble.
We can tell that the general structure of the two nebulae is the same, but that in run M3-50
the `ears' are more pronounced. Also, the higher energy in run M3-50 manages
to prevent the concentration of large amount of gas along the symmetry axis.
That the two runs have different amount of mass along the symmetry axis suggests that
this mass concentration is not a numerical effect. It is real, but its exact structure
depends on the numerics, e.g., the axisymmetric grid forces it to be exactly on the axis.
In run M50 (Fig.~\ref{M50}) the expansion velocity at $t=1140 \yr$ and at a distance of
$r=2.7 \times 10^{17} \cm$ on the symmetry axis is $v_{exp} = 58 \km \s^{-1}$,
while in run M3-50 (Fig.~\ref{M3-50}) the expansion velocity at $t=840 \yr$ and
at $r=3 \times 10^{17} \cm$ on the symmetry axis is $v_{exp} = 85 \km \s^{-1}$.

\begin{figure}
\centering

\resizebox{0.9\textwidth}{!}{\includegraphics{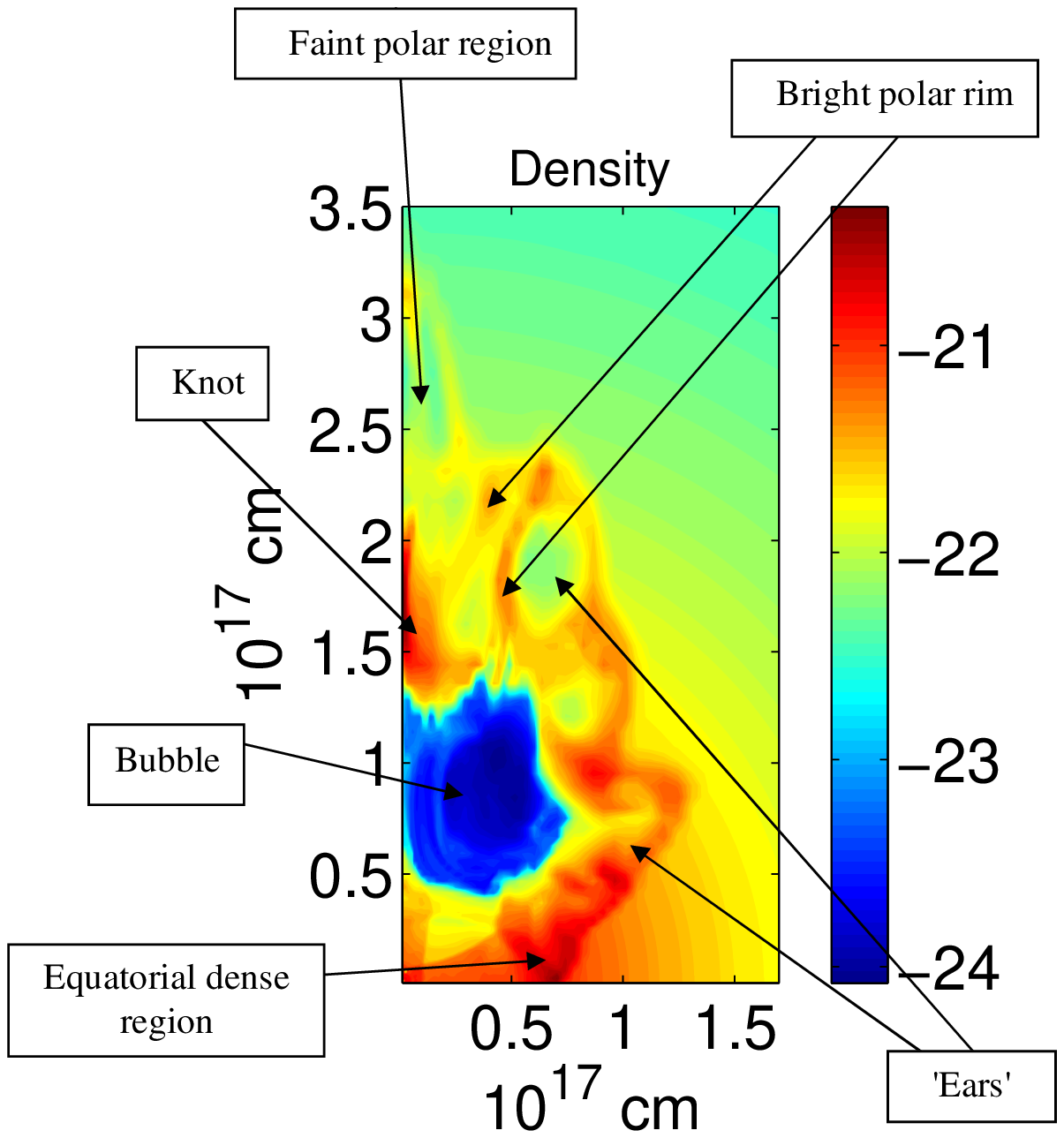}}

\caption{Density map for the M80 run at $t=1140 \yr$.
Marked are some morphological features, some of which can be
identified in the image of M1-92 (Fig.~\ref{M1-92}). }
\label{M80}
\end{figure}

These two runs demonstrate how a simple variation in the injected energy can
change not only the expansion rate, but also some morphological features.
This sensitivity is further demonstrated by run M80 that is presented at
$t=1140 \yr$ in Fig.~\ref{M80}. It is similar to run M50, but the jet was active
for a time period of $\Delta t_2= 80 \yr$ instead of $\Delta t_2= 50 \yr$ in run M50.
Noticeable is the difference in the hot bubble size obtained in runs M50 and M80.
On Fig.~\ref{M80} we try to identify some morphological features appearing in the
image of the pre-PN M1-92 (Fig.~\ref{M1-92}).
There are some quantitative differences in these features, but the general structure
of the simulations is the same as that of M1-92, beside the `ears' that envelope the
lobes in the simulations but are not observed in M1-92.
Later on we will obtain lobes with smooth boundaries, but then other features appearing
in M1-92 are less pronounced, or don't exist at all. This discrepancy will have to
be resolved when we better understand the interaction of the jets with the slow wind.
It is possible that a different AGB mass loss history will remove the ears if the
run is continued to later times.
In a future paper we will also check whether a more massive jet can form better agreement.

\begin{figure}
\centering
\resizebox{0.6\textwidth}{!}{\includegraphics{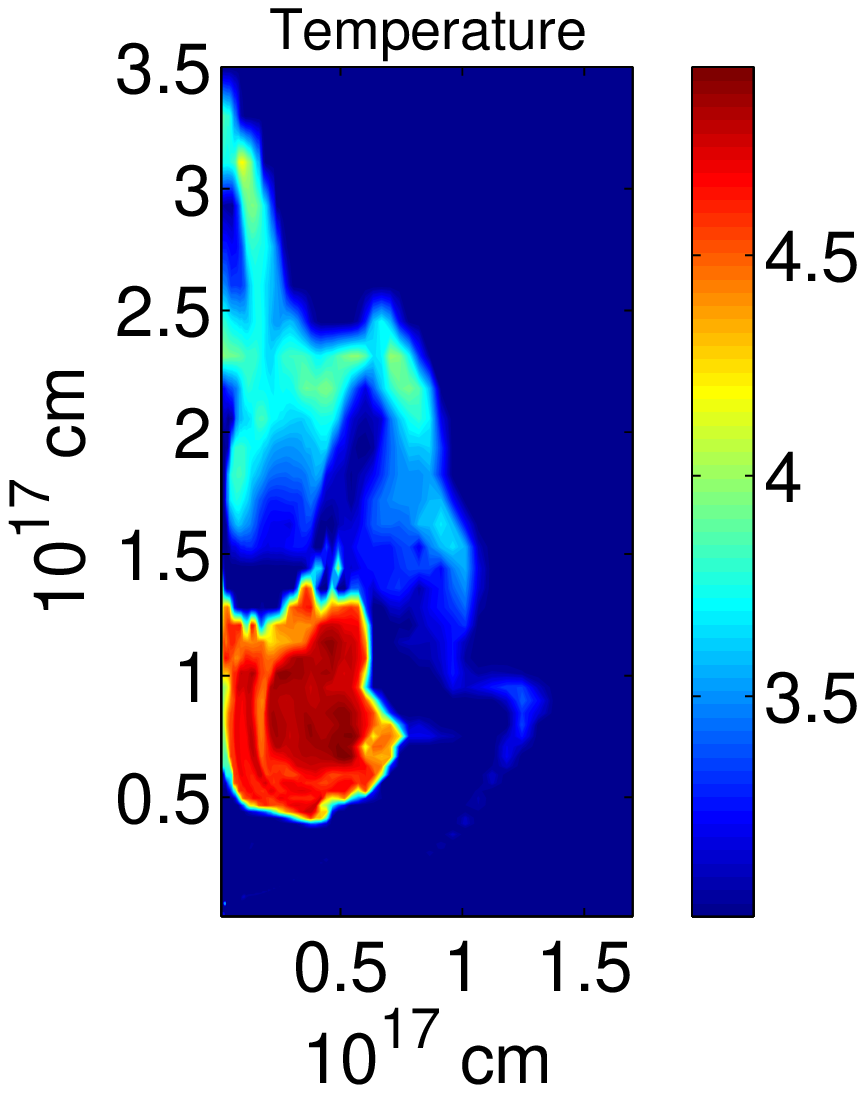}}

\caption{Temperature map of the M80 run at  $t=1140 \yr$.
The temperature scale on the right is in logarithmic units of degrees in Kelvin.}
\label{m2temp}
\end{figure}

In any case, we suggest that the basic structure of M1-92 can be formed by a jet similar
to that in run M80. The region of the knot along the symmetry axis  (or knots if there are
several knots)
closest to the center of the nebula is within the main shell.
This is best seen in Fig.~\ref{M0} (Akashi 2008).
This is a result of the usage of a wide jet that form vortices in the hot
bubble (Fig.~\ref{M0}). When the jet is narrow, no such knots are
formed (Lee \& Sahai 2003).
The inner regions of knots cool to low temperatures, but their outskirts can
become warm, $\sim 10^4 \K$ (Fig.~\ref{m2temp}), and be a source of emission lines,
such as found by Bujarrabal et al. (1998b) in the knots of M1-92 (For more on
line emission from shocked bullets see Raga et al. 2008).

We cannot predict at this point the X-ray emission of M1-92.
Examining the jet simulated here, we find that the shocked jet's material had
time to cool (both radiative and adiabatic cooling are included in our code) to
a temperature of $T \ll 10^6 \K$.
This is seen in the temperature map given in Fig.~\ref{m2temp}.
This implies that the shaping jet will not be a source of X-ray emission.
However, a second jet-launching episode, or a spherical wind from the central
star, can fill the cavity with a hot enough gas to be a strong source of X-ray emission.

\begin{figure}
\resizebox{0.5\textwidth}{!}{\includegraphics{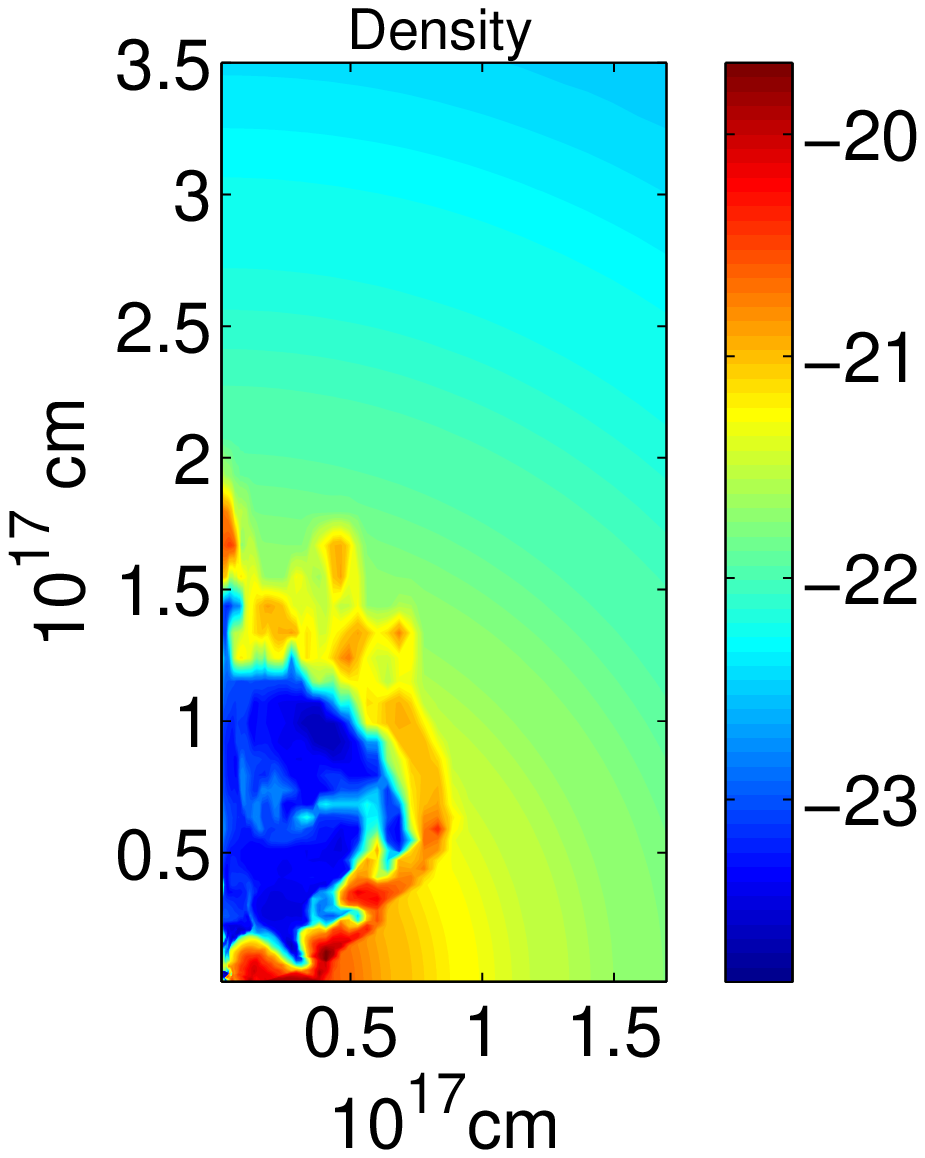}}
\hskip -0.5in
\resizebox{0.5\textwidth}{!}{\includegraphics{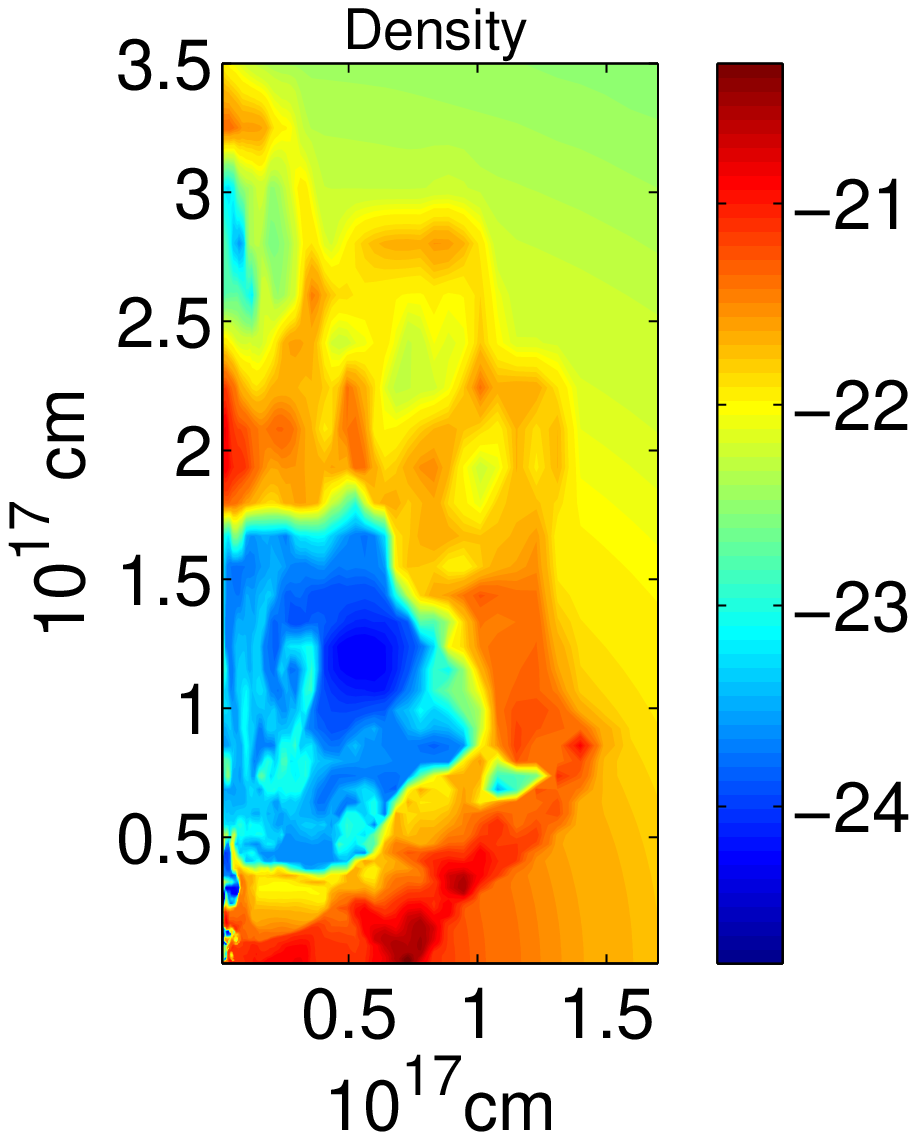}}

\caption{Density map for the M150 run, at $t=540 \yr$ (left) and $1140 \yr$ (right). }
\label{M150}
\end{figure}

\begin{figure}
\centering
\resizebox{0.8\textwidth}{!}{\includegraphics{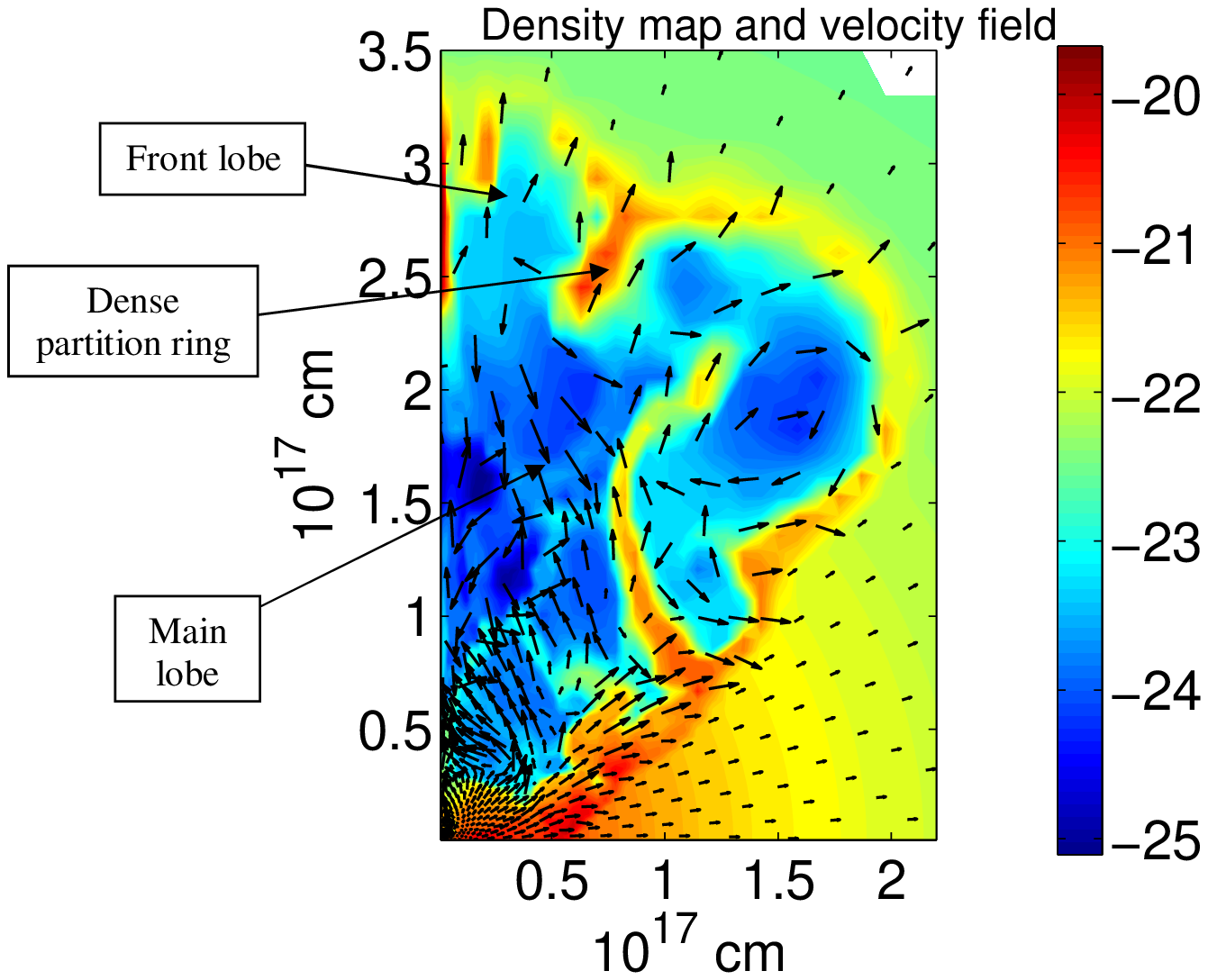}}
\caption{Density map for the M600 run at $t=840 \yr$.
Marked are features that can be identified in the image of Mz~3 (Fig.~\ref{Mz3}).
Arrows indicate flow direction, with a scale of: $v > 200 \km \s^{-1}$ (long arrow),
$ 20 < v \le 200 \km \s^{-1}$ (medium arrows),
and $ v \le 20 \km \s^{-1}$ (short arrows).}
\label{M600}
\end{figure}

\begin{figure}
\centering
\resizebox{0.7\textwidth}{!}{\includegraphics{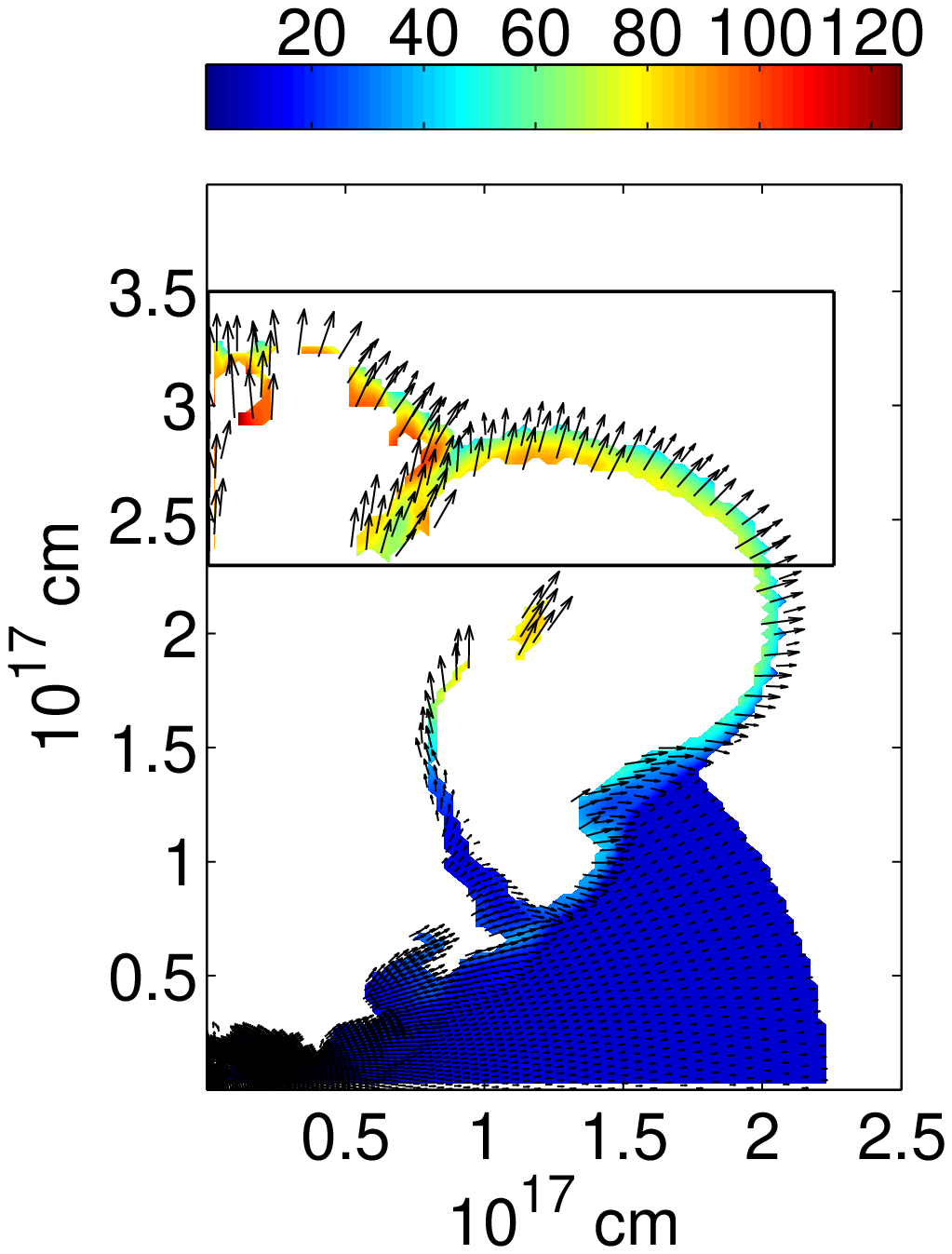}}
\caption{Velocity map for regions having a density of $\rho>10^{-22} \g \cm^{-3}$ for
the M600 run at $t = 840 \yr$.
The magnitude of the velocity is represented by the color, with the scale as given by the
bar in units of $\km \s^{-1}$.  The arrows give only the direction.
The rectangle show the region termed `cap'. }
\label{m600v}
\end{figure}

To close the study of the sensitivity to the duration of the jet active phase,
we show two more cases in Figs.~\ref{M150}-\ref{m600v}. In Fig.~\ref{M150} we present
run M150, where beside a vortex around the lower density region of the hot
bubble, the nebular gas approaches a ballistic motion. In Fig.~\ref{M600}
the jet duration is relatively long, and a large bubble with a complicated flow
inside it is formed.

The two lobes of Mz~3 might have been formed by two jet-launching
episodes (in addition to the mass loss episodes that formed the
structures outside the lobes), and not by a single episode as we
examine here. There are some arguments suggesting that two shaping
episodes of the lobes and front lobes is not unlikely. There have
been episodic bipolar ejections before the one or two episodes
that formed the lobes and front-lobes (Guerrero et al. 2004). The
X-ray image of Mz~3 suggests the presence of a relatively narrow
jet inside the main lobes, fitting the size of the front lobe
(Kastner et al. 2003). The long-slit spectrum of Mz~3 reported by
Santander-Garcia et al. (2004) shows that the dense partition
ring$-$on the boundary of the main lobe and the front lobe$-$has a
broader spectrum. This spectrum suggest that this region is moving
faster, or that there is a turbulence there.

In any case, here we point that a single jet-launching episode can
form a front-lobe type of structure. We marked some morphological
features in Fig.~\ref{M600}, that resemble to some degree
morphological features observed in Mz~3 as marked in Fig.
\ref{Mz3}. The run presented in Fig.~\ref{M0} (Akashi 2008) better
represent the morphology of Mz~3. As stated earlier, our goal is
not to fit specific PNs, but rather to point to the general
morphological features that jets can form. In the case of Mz~3 the
task is more difficult because the two lobes are not exactly
identical, in particular in the front lobe region. We see by
comparing figures  \ref{Mz3} and \ref{M600} that the match is
definitely not perfect, but when the results presented
Fig.~\ref{M0} are considered as well, we can conclude that jets
can indeed form the same type of morphological features as
observed in Mz~3, with different parameters for the two lobes. The
lobe appearing in the right side of Fig.~\ref{Mz3} is better
reproduced by the M150 run shown in Fig. \ref{M150} (right panel).
The `ears' on the front lobe are clearly seen in this numerical
run.  Also seen is the relatively narrow waist in the equatorial
plane with a dense ring (torus) at a distance of $\sim 6 \times
10^{16} \cm$ from the center.

As seen in Fig.~\ref{M600}, the flow structure of the gas in the hot low-density bubble,
formed by the shocked jet material, is quite complicated. For example, we can see a vortex in the
right side of the bubble, and material flowing inward$-$a backflow.
The backflow results from the decrease of the pressure in the inner region, which itself result from
the expansion of the nebula as it follows the dense shell.
In Fig.~\ref{m600v} we study the expansion of the dense regions, that later form the bright
regions of the nebula observed in the visible band.
We examine only regions having densities of $\rho>10^{-22} \g \cm^{-3}$, and show the
magnitude of the velocity by a color coding, as indicated in the
figure, and the direction by arrows.
The front of the shell is termed `cap' (Balick 2004) as marked on Fig.~\ref{m600v}.
As evident from the figure, different regions in the `cap' have different velocities that are
typically much larger than in the rest of the shell.
The cap region of Mz~3 also show large velocities (Guerrero et al. 2004). In the case of Mz~3
some of the fast moving regions might be outside the main lobe (Guerrero et al. 2004), and
are not studied here. However, our results show that the `cap' itself can have relatively large
velocities.

\subsection{Dependance on the Slow Wind Density Profile}
\label{sec:prof}

\begin{figure}
\centering
\resizebox{0.6\textwidth}{!}{\includegraphics{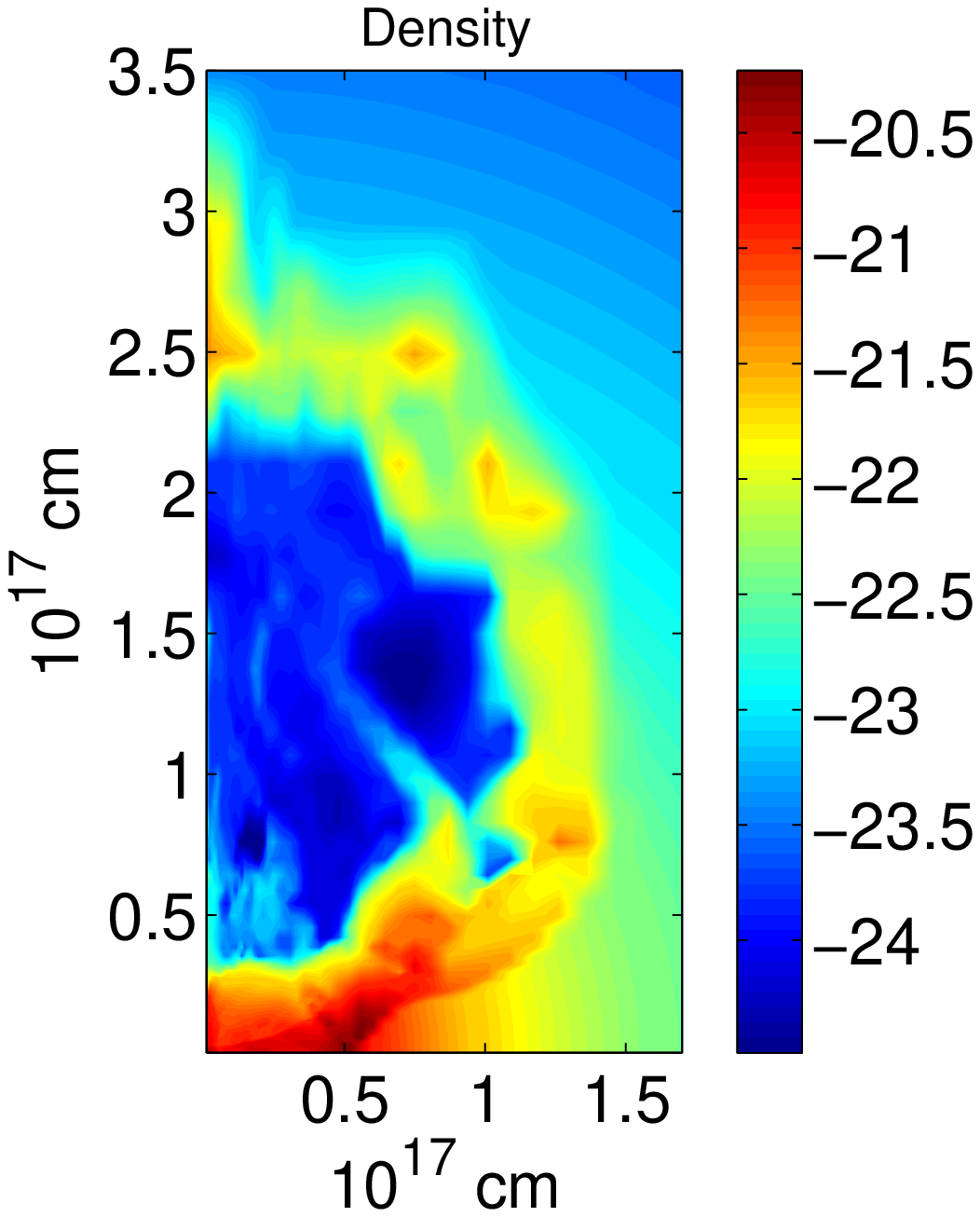}}

\caption{Density map for the M3 run at $t= 840 \yr$. }
\label{M3}
\end{figure}

\begin{figure}
\centering
\resizebox{0.7\textwidth}{!}{\includegraphics{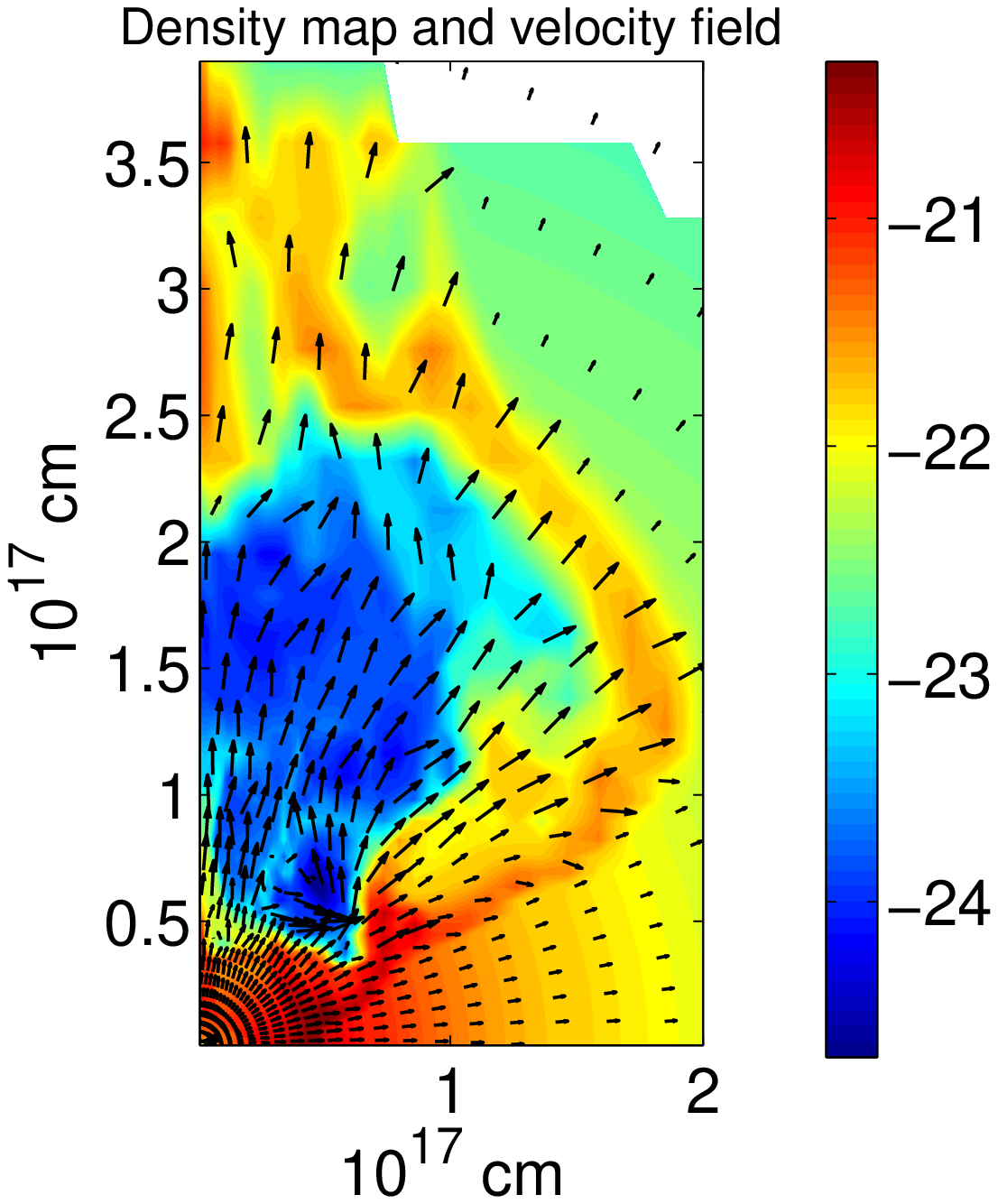}}

\caption{Density map for the M2.2-3 run at $t=840 \yr$.
Arrows indicate flow direction, with a scale of: $v > 200 \km \s^{-1}$ (long arrow),
$ 20 < v \le 200 \km \s^{-1}$ (medium arrows),
and $ v \le 20 \km \s^{-1}$ (short arrows). }
\label{M2.2-3}
\end{figure}

In Fig.~\ref{M3} and Fig.~\ref{M2.2-3} we present the results for runs where the
mass loss rate of the slow wind that formed the nebula was not constant.
The initial density profile was according to equation (\ref{dens}) with
the value of $\beta$, and other parameters, given in Table 1.

Two interesting features are seen.
($i$) Comparing run M3 and M2.2-3 with run M50 (Fig.~\ref{M50}), we see
that the sides of the lobes are smoother, i.e., the `ears' are less pronounced.
($ii$) A front lobe is seen in both runs M3 and M2.2-3. However, the front lobe in run M2.2-3 is
disrupted, with `ears' type structure. The type of structure obtained in run M2.2-3
is similar in some respects to the disrupted front lobe seen in one of the
lobes of Mz~3 (at the right side of Fig.~\ref{Mz3}).

\subsection{The formation of a torus (ring)}
\label {sec:formation}

\begin{figure}
\resizebox{0.65\textwidth}{!}{\includegraphics{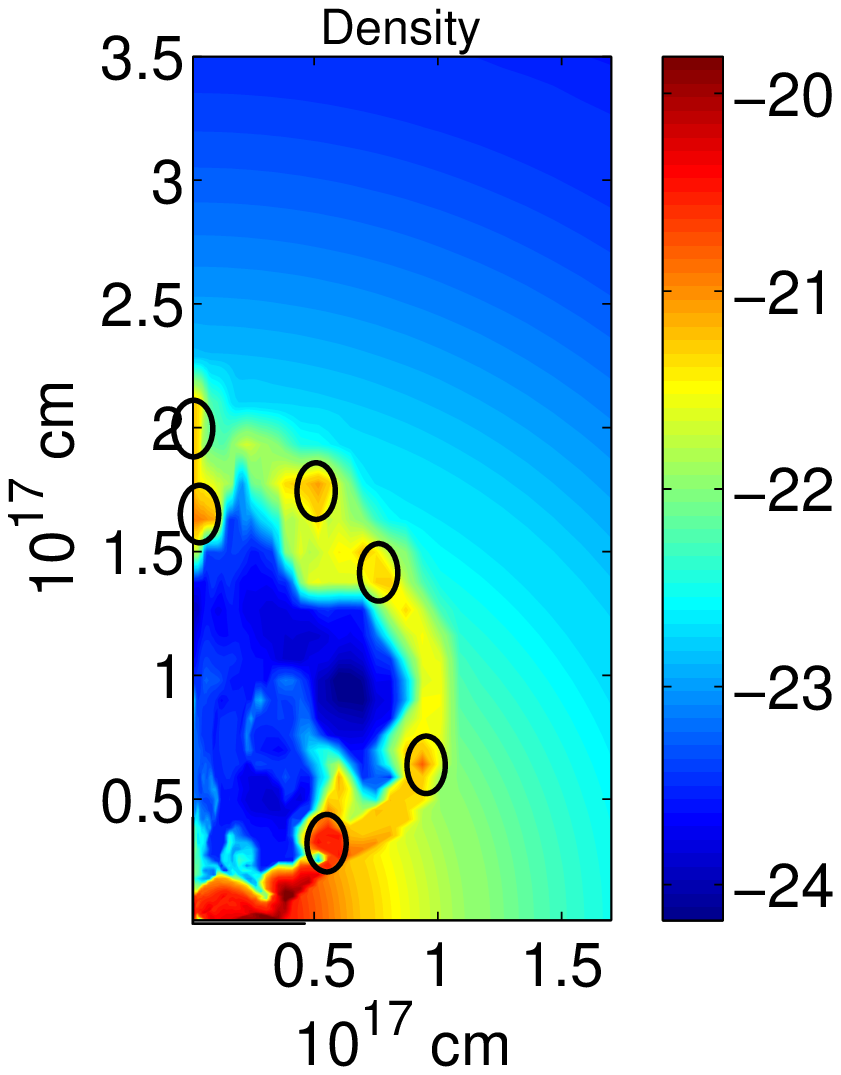}}
\hskip -1in
\resizebox{0.65\textwidth}{!}{\includegraphics{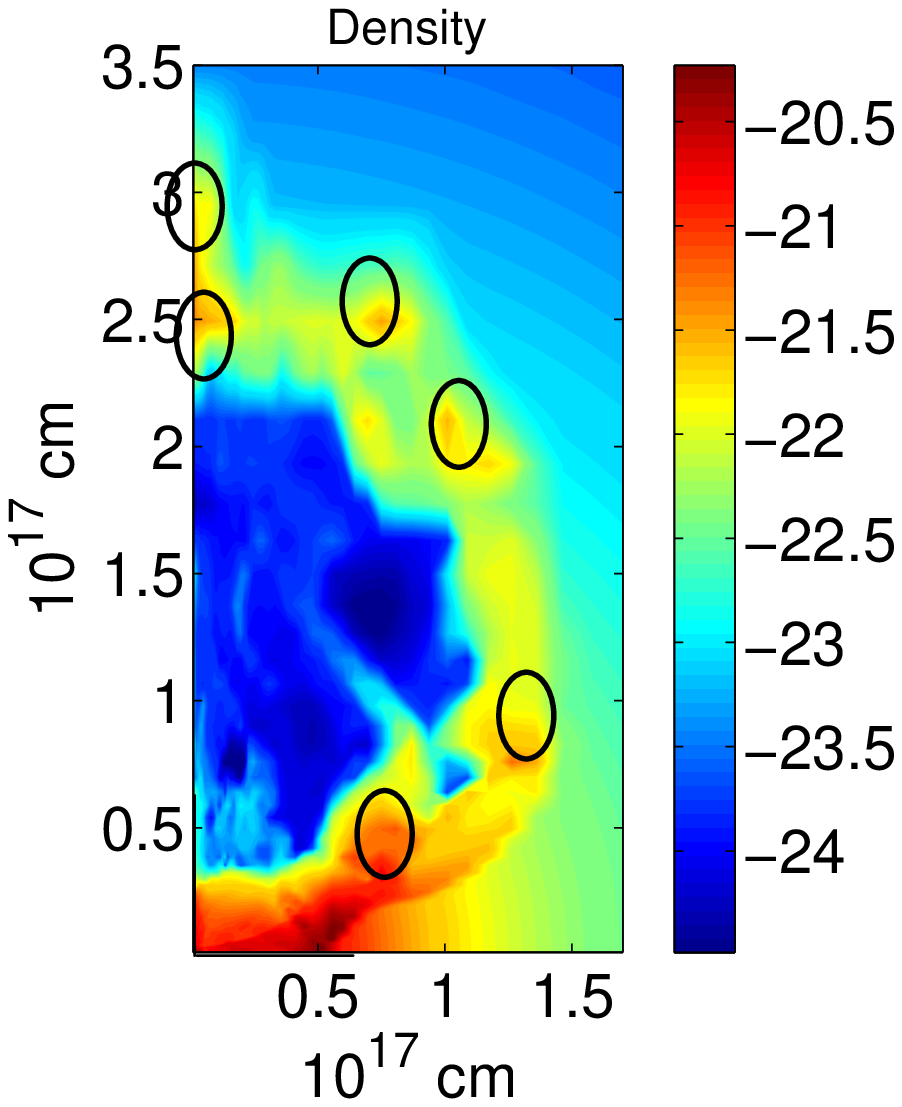}}
\caption{Density map for the M3 run at  $t=540 \yr$ (left) and
$t=840 \yr$ (right). Several dense regions are marked on the left panel. The distances
of the marked regions are increased by the same factor, and copied to the right panel
to match the location of the jet. The other regions `lag' behind, with differences
increasing toward the equator. Namely, the jet is moving relatively faster.
Dividing the distance from the center by velocity for the different regions might
give the wrong impression that the jet was launched after regions closer to the equatorial plane. }
\label{compare}
\end{figure}

A dense, slowly expanding ring (torus) is formed in most runs presented in this paper and
in Akashi (2008). This is an important result proposed by Soker \& Rappaport (2000).
Such rings are observed in many PNs.
Huggins (2007) studied several PNs, and found that the distance $r$ divided by expansion
velocity $v_r$, $\tau = r/v_r$, of the jets and the torus in each PN give somewhat
different values, with lower values for jets.
Still, their age is not much different and Huggins (2007) concluded that the
formation of the jets and the formation of the torus are connected.
He concluded that the torus (ring) is probably ejected up to several hundreds years before
the jets.
We show that this is not necessarily so.
In Fig.~\ref{compare} we show run M3 at two times. We mark several features at $t=540 \yr$
(left panel). We then increase the distances of all the marked regions from
the center by the same factor, and mark the results on the density image at $t=840 \yr$
(right panel), such that the marks on the two dense parts along the symmetry axis (the `jet')
coincide.
We clearly see that as we move from axis toward the equator the dense regions lag behind
more and more. Namely, it looks as if the jets expands faster than the other regions.
The distance from the center divided by the difference is smaller for the jet
than for other regions, despite that all features were formed by the same event.
This behavior result from the fast decline in the density, $\beta > 2$; performing the
same exercise for run M50, where $\beta=2$, leads to less conclusive results.
Namely, the equatorial flow suffers more deceleration than the polar flow, therefore,
the expansion speed history of different regions is not the same.
We did not compare the equator itself for a numerical reason. We inject a small amount
of material in the equator during the jet launching phase for numerical reason. This
non-physical material accelerate somewhat the ring (torus).

\begin{figure}
\centering
\resizebox{0.6\textwidth}{!}{\includegraphics{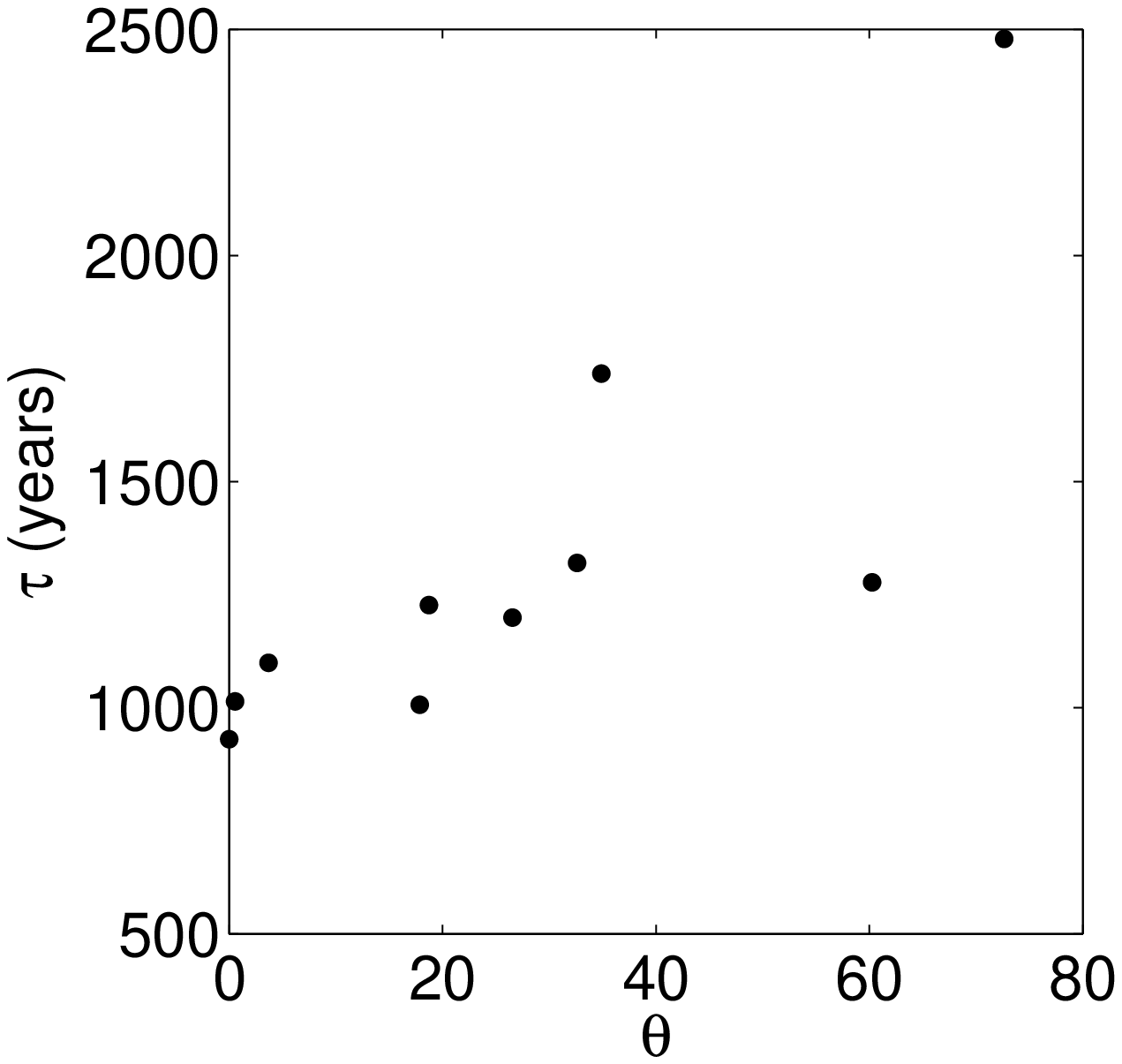}}
\caption{The values of $\tau=r/v_r$ for the regions marked in Fig.~\ref{compare}
and few other dense regions at a later time of $t=960 \yr$.
Here $v_r$ is the radial velocity of the region, and $r$ its distance from the center.
$\theta$ is the direction of the dense region measured from the symmetry axis; $\theta=0$
is the jet direction. Although all features where formed by the same mass loss episode,
it appears as if the jet is younger than features formed closer to the equator
(at $\theta=90^\circ$). }
\label{tau10}
\end{figure}

To further demonstrate this result, we calculate the values of $\tau=r/v_r$ for the regions marked in
Fig.~\ref{compare} and few other dense regions, at a later time of $t=960 \yr$.
In Fig.~\ref{tau10} we show the values of $\tau$ calculated in these regions
as function of the angle from the symmetry axis. Observational study as that of
Huggins (2007) will identify the features at $\theta=0^\circ$ as the jet, and those
near $\theta=90^\circ$ as the torus.
It is clear from this figure that the jet appears as if it was formed after the torus.
However, they were all formed from the same mass loss episode.
We note that Alcolea et al. (2007, 2008) find the age of M1-92 to be $\sim 1200 \yr$, and claim that
it was formed by a single shaping event that last for $<100 \yr$.

In a test run (not shown here) we increased the mass loss rate of both the
slow wind and the jet by a factor of 50. The other parameters were as in run M80.
About half of the mass of the nebula was concentrated in the torus.
We find that we can account for cases where the mass in the torus is about equal to the
mass in the rest of the nebula.
This leaves one problem that may still exist with our interpretation of the results of
Huggins (2007), and this is, as noted by Huggins (2007), the large mass of the torii
in some cases.
In cases where the torus mass is much more than the mass in the rest of the nebula
our model cannot account for the whole torus, and a special mass loss episode must
be added to the model, e.g., by a common envelope interaction.

\subsection{The linear $v-r$ relation of the flow}

\begin{figure}
\centering
\resizebox{0.4\textwidth}{!}{\includegraphics{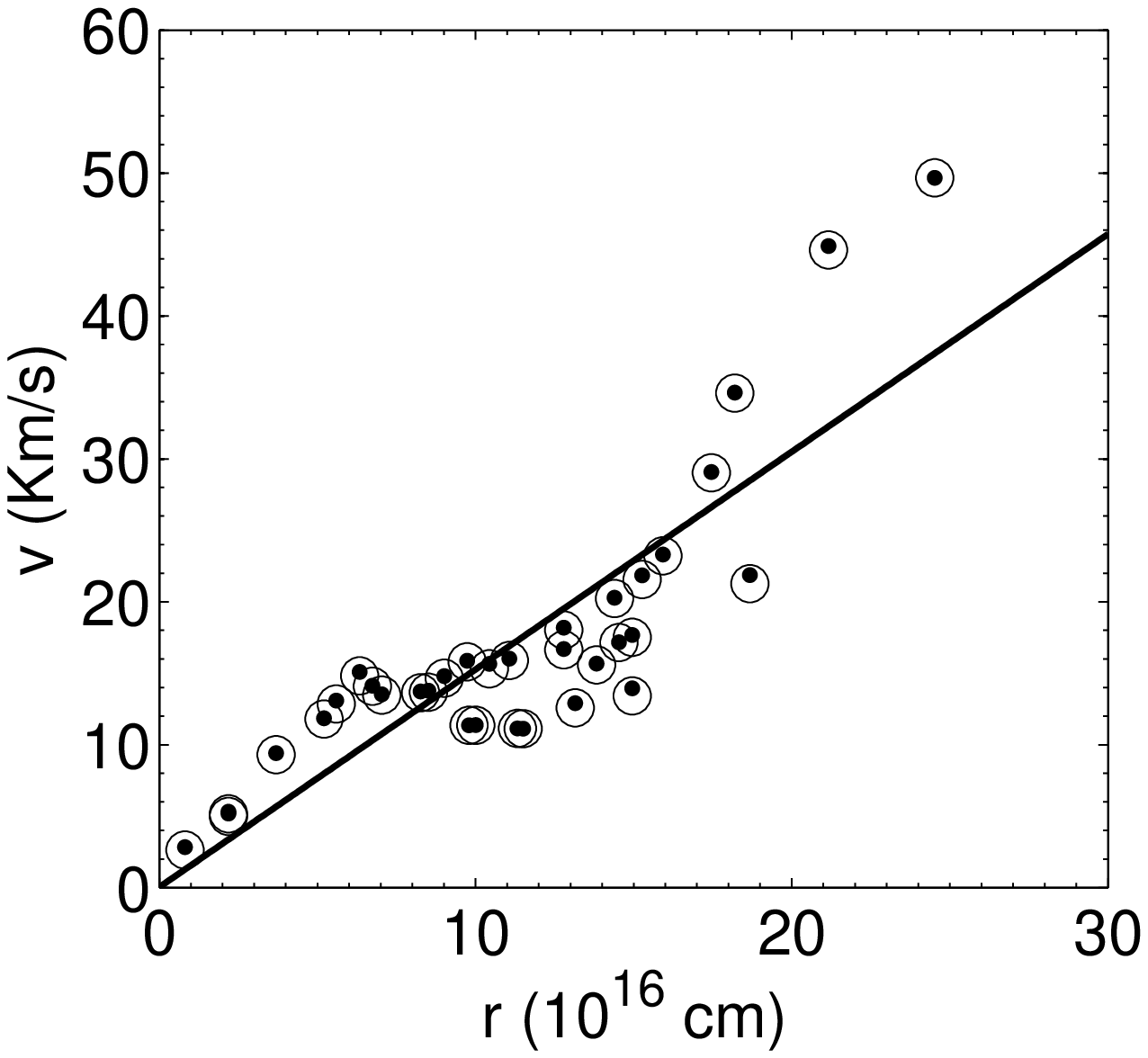}}
\resizebox{0.4\textwidth}{!}{\includegraphics{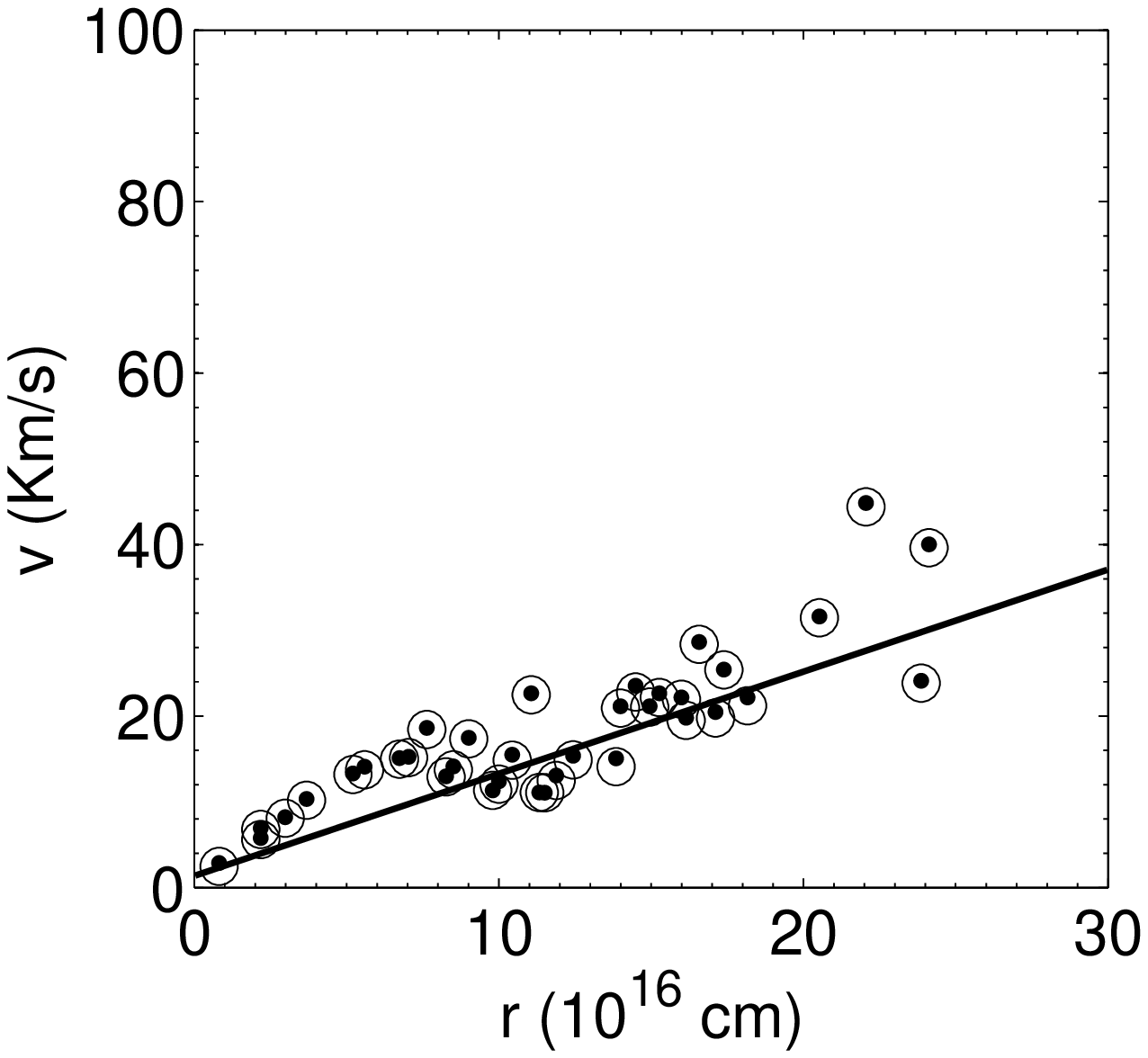}}
\caption{Radial velocity (circles) and total velocity (dots) versus
distance from the center of different small regions in the dense
nebula for the M80 (left) and M2.2 (right) runs.  }
\label{linflow1}
\end{figure}

As mentioned in section \ref{sec:intro} in many PNs there is a linear relation between distance
from the center and velocity.
To check this relation for our simulations, we calculate the radial velocity for
many small regions in the nebular shell.
The radial velocity of each region was taken to be the center of mass velocity of the region.
In Fig.~\ref{linflow1} we draw the radial velocity as function of distance for such regions
in the M80 and M2.2 runs. The M2.2 run has no density map in the paper. It has an initial shallower
slow wind density profile than run M3, but other than that has all parameters as in run M3.
We fit these scattering plots with a linear line.
A nice linear relation is seen in run M2.2.
Although not perfect in run M80, a linear relation holds there as well.

\begin{figure}
\centering
\resizebox{0.4\textwidth}{!}{\includegraphics{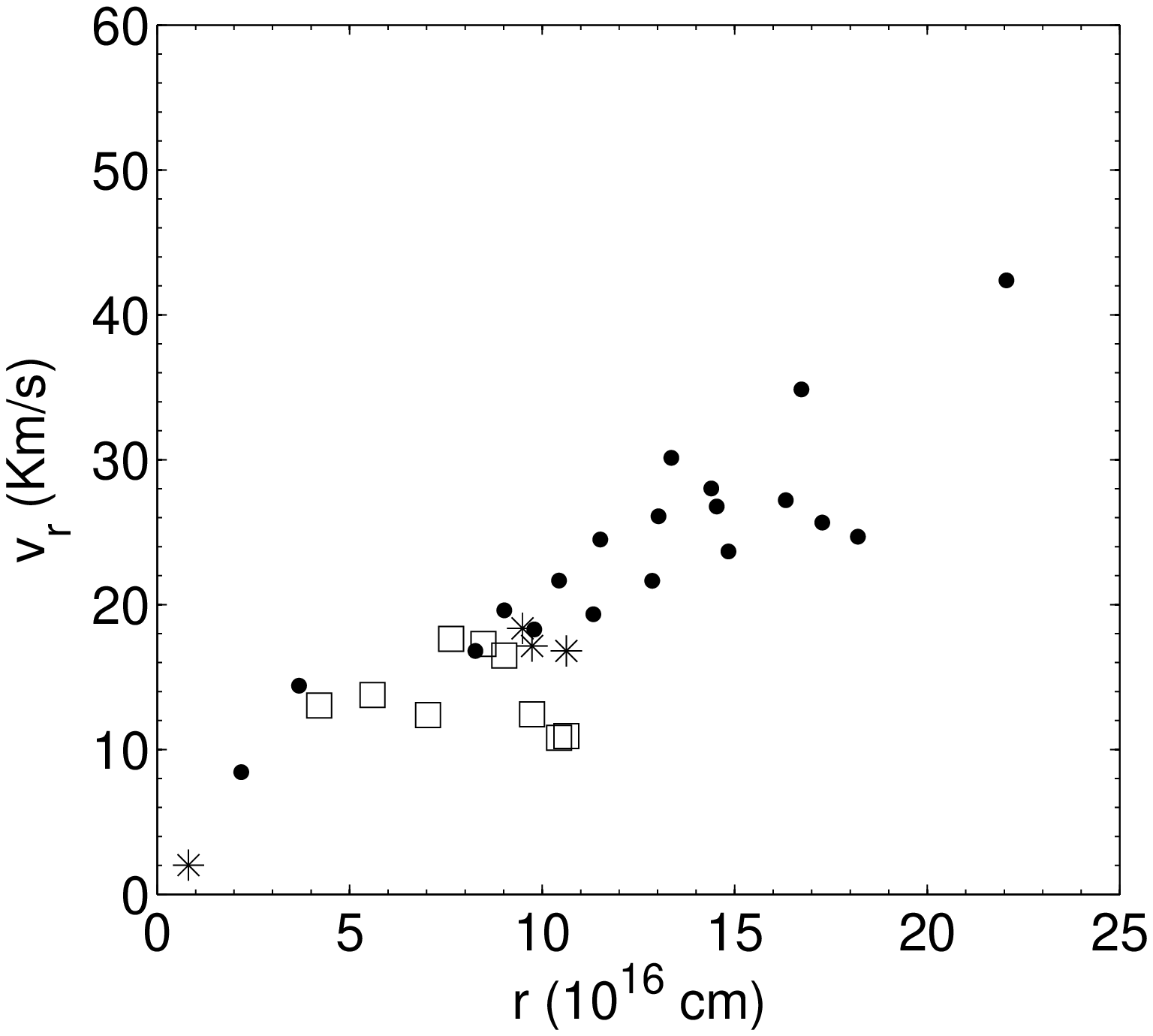}}
\resizebox{0.4\textwidth}{!}{\includegraphics{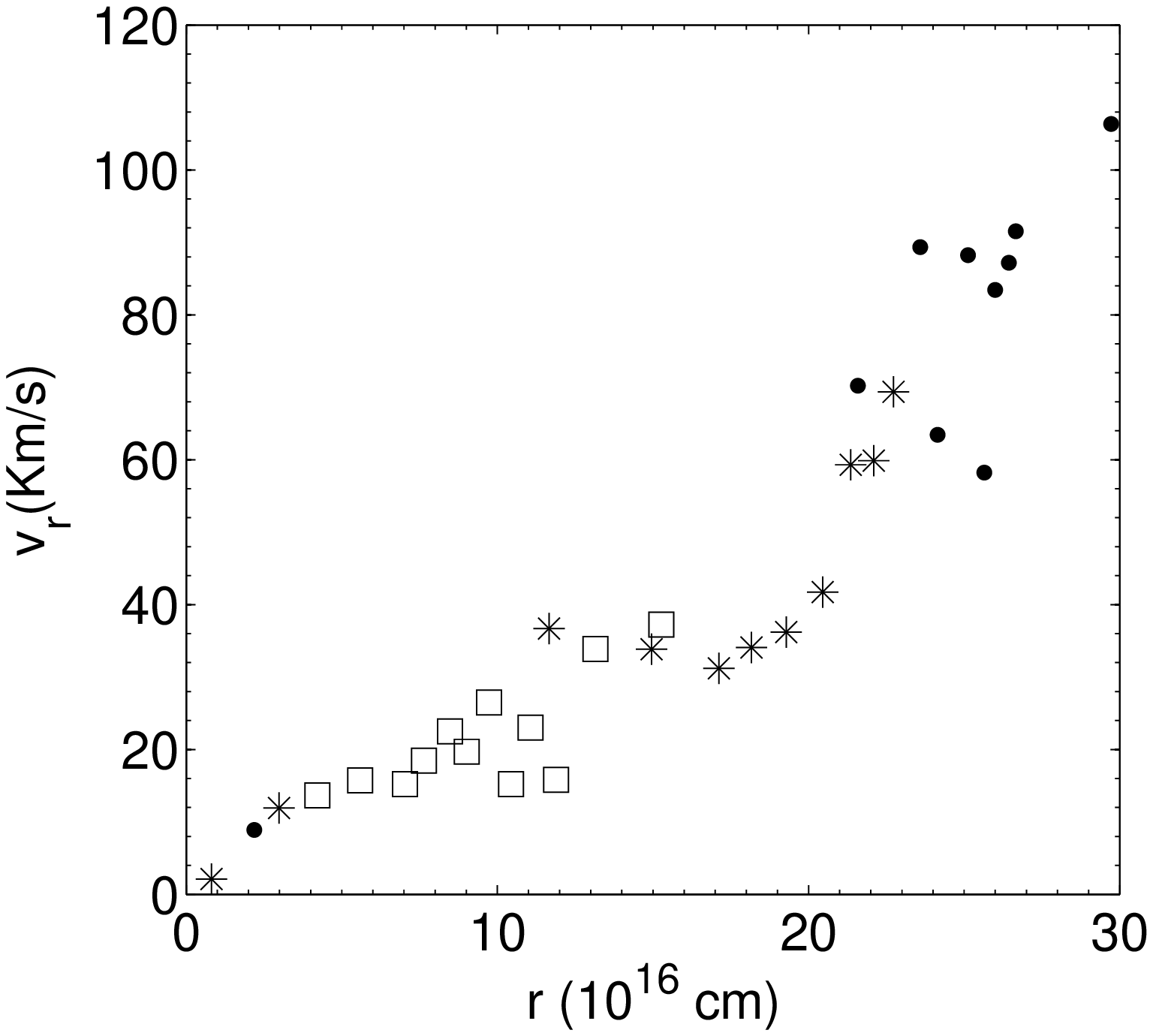}}
\caption{Radial velocity versus distance from the center of different
small regions in the dense nebula for the M50 (left) and M3 (right) runs.
Dots marked region that are within $25^\circ$ from the symmetry axis (vertical axis),
while asterisks and squares mark regions within a direction $25^\circ-50^\circ$ and $50^\circ-75^\circ$
from the symmetry axis, respectively. }
\label{linflow2}
\end{figure}

For run M50 and M3 we make the same plots, but mark differently regions according to their direction from the
symmetry axis, as indicated in the figure caption of Fig.~\ref{linflow2}.
In the case of run M50 a general linear relation holds for all regions.
In the case of run M3, where the slow wind density decline steeply with radius, regions closer to the
symmetry axis reach higher velocities, as was discussed in section \ref{sec:formation}.
This is clearly see in Fig.~\ref{linflow2}. Here again the observational results of Huggins (2007)
is accounted for by one shaping episode.
Although there is only one shaping episode, i.e., the jet, this explanation requires that
the spherical slow wind mass loss rate will increase hundreds of years prior to the
jet ejection.

We conclude that the type of interaction studied here can explain the linear relation observed in many PNs,
as well as the seemingly result that jets are younger than torii.

\section{SUMMARY}

We conducted numerical simulations of axisymmetrical jets expanding into a spherical
AGB slow wind. Using the axisymmetrical nature of the flow and its mirror symmetry about
the equatorial plane, we simulated the 3D flow with a 2D hydrodynamical numerical code
covering one quarter of the meridional plane. The parameters of the slow wind and jets
are summarized in Table 1.
In the present study we have concentrated on the types of structures represented by the
bipolar PN Mz~3 and the bipolar pre-PN M1-92.
We used light jets (CFW), i.e., their density is much lower than the
slow wind density at the same radius.
Our goal was not to play with the different parameters
in order to fit as best as possible each structure, but rather to show that the basic types
of structures can be reproduced, and to study the role of some parameters.
As indicated in the caption of Table 1, many parameters were held fixed in the present study.

We did not vary the slow wind mass loss rate at the end of the AGB, $\dot M_1=10^{-6} M_\odot \yr^{-1}$,
that appears in the density profile (eq.~\ref{dens}).
By scaling the mass loss rate in the slow wind and the jet one can produce more massive nebulae.
In a test run (not shown here) we increased the mass loss rate of of both the
slow wind and the jet by a factor of 50, keeping the other parameters as in run M80.
We obtained a nebula similar in structure to that in run M80, but typical densities
larger by a factor of $\sim 50$ in the different regions, and much lower temperatures.
The main differences in the structure were the structure of the `ears', as they are
the results of instabilities that are sensitive to other small vatiations.
These differences are the result of the shorter radiative cooling time of the denser gas.

Our main results are summarized below. One should remember that the axisymmetrical numerical
grid forces some features, in particular dense clumps, to be formed exactly along the symmetry
axis. These are real features, but in reality such features can be off-axis and more
dispersed.
\begin{enumerate}
\item With a single episode of jet's (CFW) launching we could reproduce a lobe structure
having a `front-lobe', as indicated in Figs.~\ref{M600} and \ref{M0}; the run represented
in Fig.~\ref{M0} is from Akashi (2008), and does not appear in Table 1. the `front-lobe' is
similar to the one seen on one of the lobes of Mz~3, as indicated in Fig.~\ref{Mz3} (left lobe).
The other lobe of Mz~3, on right side of Fig.~\ref{Mz3}, has a structure that was reproduced
by the run shown in Fig.~\ref{M150}, and to some degree by that shown in Fig.~\ref{M2.2-3}.
We cannot rule out the possibility that the front lobe was formed by a separate
jet; we only showed that it can be reproduced by a single jet-launching episode.
\item In some runs (e.g., Fig.~\ref{M80}) dense clumps are formed along the symmetry axis, such as those observed in the pre-PN M1-92 (Fig.~\ref{M1-92}).
\item Instabilities lead to the formation of `ears' on the lobe (bubble) boundary,
e.g., Fig.~\ref{M80}. Such `ears' are observed in the front-lobe of one of the two lobes of
Mz~3 (on the right side of Fig.~\ref{Mz3}), but in many other PNs such ears are not seen.
\item We run several cases where the density of the slow wind declines faster
than $r^{-2}$, i.e., $\beta>2$ in equation (\ref{dens}), as is expected when the mass loss rate
increases at the end of the AGB.
In those cases (Figs.~\ref{M3}, \ref{M2.2-3}) the `ears' disappear from most
of the lobe boundaries, and might appear only in the front.
In particular, in Fig.~\ref{M2.2-3} we can see a structure where only the front lobe is
disrupted by `ears'.
\item The different runs performed here and in Akashi (2008) show that the mass loss history
of the slow wind has a profound influence on the PN structure. Relatively small
variations in the slow wind mass loss history can lead to new morphological features.
\item A dense expanding torus (ring; disk) is formed in most of our runs,
as proposed by Soker \& Rappaport (2000).
It is likely that in some types of binary interaction an expanding torus is formed as well.
Our results show that a ring (torus) is easily formed by jet shaping, with no need for a specific
equatorial mass loss process. However, if the torus is very massive, much more than half
the total mass in the nebula, then a special equatorial
mass ejection process should be added to the model (Huggins 2007).
\item The torus and lobes are formed at the same time and from the same mass loss
rate episode.
However, in runs with $\beta>2$ the ratio of the distance divided by the radial velocity,
$\tau=r/v_r$, is larger for regions closer to the equatorial plane
(fig.~\ref{compare}, \ref{tau10}, \ref{linflow2}).
If the ratio $\tau=r/v_r$ is taken to be the age of the specific regions, then
one might deduce that the torus was formed before the jets, although this is not the case:
they were all formed from the same shaping mass loss episode.
\item Our results, as depicted in Fig.~\ref{linflow1} and  \ref{linflow2} show that
impulsive jets, i.e., a short active phase, lead to a linear relation (wrongly termed
by some authors as a Hubble flow) between distance and expansion velocity.
\item Dense regions in the shell reach sufficiently high velocities to shock-excite
visible emission lines (Fig.~\ref{m600v}).
Indeed, some dense regions reach temperatures of $\ga 10^4 \K$ (Fig.~\ref{m2temp}).
\end{enumerate}

\acknowledgments

We thank Bruce Balick, Patrick Huggins, and Raghvendra Sahai for many clarifying
and useful comments.
This research was supported in part by the Asher Fund for Space Research at the
Technion.

\end{document}